%% file: main.tex
\documentclass[12pt,superscriptaddress,showpacs,nobibnotes,longbibliography]{iopart}

\usepackage[margin=1in]{geometry}
\usepackage[colorlinks,urlcolor=blue,citecolor=blue,linkcolor=blue]{hyperref}
\usepackage[pdftex]{graphicx}
\usepackage[usenames,dvipsnames]{xcolor}
\usepackage{url}
\usepackage{tabularx}
\usepackage{tabulary}
\usepackage{multirow} 
\usepackage{multicol}
\usepackage{graphicx}

\expandafter\let\csname equation*\endcsname\relax
\expandafter\let\csname endequation*\endcsname\relax
\usepackage{amssymb,amsmath}

\usepackage{pdfpages}
\usepackage{abstract}
\usepackage{wrapfig}
\usepackage{fancyhdr}
\usepackage{parskip}
\usepackage{color}
\usepackage{enumitem}
\usepackage{bm}
\usepackage{lipsum}
\usepackage[sc]{mathpazo}
\usepackage{subcaption}
\usepackage{titlesec}
\usepackage{lineno}
\usepackage{nicefrac}
\usepackage[titletoc]{appendix}
\usepackage[sort&compress,numbers]{natbib}
\usepackage{doi}
\usepackage{nameref}
\usepackage{comment}
\usepackage[draft, inline, nomargin]{fixme}

\usepackage[normalem]{ulem}

\fxsetup{theme=color, mode=multiuser}
\FXRegisterAuthor{rc}{1}{\color{blue}RC}
\FXRegisterAuthor{bl}{2}{\color{red}BL}
\FXRegisterAuthor{pm}{3}{\color{purple}PM}

\newcommand{\nuebar}{\ensuremath{\overline{\nu}_{e}} }
\newcommand{\uFive}{$^{235}$U}
\newcommand{\uEight}{$^{238}$U}
\newcommand{\pNine}{$^{239}$Pu}
\newcommand{\pOne}{$^{241}$Pu}
\newcommand{\Pone}{PROSPECT-I}
\newcommand{\Ptwo}{PROSPECT-II}

\begin{document}

\title[]{PROSPECT-II Physics Opportunities}
\vspace{-10pt}
\input{AuthorListJun2021}

\pagebreak

\begin{abstract}
\input{abstract}
\end{abstract}

%
%
%
%
%

\clearpage
\tableofcontents{}
\clearpage
\markboth{}{}

\input{Introduction.tex}

\section{PROSPECT-II physics goals and discovery potential}
\label{sec:PII_PhysicsGoals}
\input{PII_PhysicsGoals.tex}

\section{PROSPECT-II detector}
\subsection{PROSPECT-I detector overview}
\label{sec:PI_Detector}
\input{PI_Detector.tex}
\subsection{Upgrades for PROSPECT-II detector}
\label{sec:PII_Detector}
\input{PII_Detector.tex}

\section{Conclusions}
\label{sec:Conclusions}
\input{Conclusions.tex}

\clearpage

\bibliographystyle{iopart-num-long}
\bibliography{main}

\end{document}

%% file: AuthorListJun2021.tex

\author{M~Andriamirado$^{ 6}$,
A~B~Balantekin$^{ 9}$,
H~R~Band$^{ 17}$,
C~D~Bass$^{ 7}$,
D~E~Bergeron$^{ 14}$,
N~S~Bowden$^{ 15}$,
C~D~Bryan$^{ 12}$,
R~Carr$^{ 10}$, 
T~Classen$^{ 15}$,
A~J~Conant$^{ 12}$,
G~Deichert$^{ 12}$,
A~Delgado$^{ 16,3}$,
M~V~Diwan$^{ 1}$,
M~J~Dolinski$^{ 5}$,
A~Erickson$^{ 11}$,
B~T~Foust$^{ 17}$,
J~K~Gaison$^{ 17}$,
A~Galindo-Uribari$^{ 16,3}$,
C~E~Gilbert$^{ 16,3}$,
C~Grant$^{ 4}$, 
S~Hans$^{ 1}$,
A~B~Hansell$^{ 8}$,
K~M~Heeger$^{ 17}$,
B~Heffron$^{ 16,3}$,
D~E~Jaffe$^{ 1}$,
S~Jayakumar$^{ 5}$,
X~Ji$^{ 1}$,
D~C~Jones$^{ 8}$, 
J~Koblanski$^{ 2}$,
P~Kunkle$^{ 4}$, 
O~Kyzylova$^{ 5}$, 
C~E~Lane$^{ 5}$,
T~J~Langford$^{ 17}$,
J~LaRosa$^{ 14}$, 
B~R~Littlejohn$^{ 6}$,
X~Lu$^{ 16,3}$, 
J~Maricic$^{ 2}$,
M~P~Mendenhall$^{ 15}$,
A~M~Meyer$^{ 2}$, 
R~Milincic$^{ 2}$, 
P~E~Mueller$^{ 16}$,
H~P~Mumm$^{ 14}$,
J~Napolitano$^{ 8}$,
R~Neilson$^{ 5}$,
J~A~Nikkel$^{ 17}$, 
S~Nour$^{ 14}$, 
J~L~Palomino$^{ 6}$, 
D~A~Pushin$^{ 13}$,
X~Qian$^{ 1}$,
R~Rosero$^{ 1}$,
M~Searles$^{ 12}$,
P~T~Surukuchi$^{ 17}$,
M~A~Tyra$^{ 14}$, 
R~L~Varner$^{ 16}$,
D~Venegas-Vargas$^{ 16,3}$, 
P~B~Weatherly$^{ 5}$, 
C~White$^{ 6}$, 
J~Wilhelmi$^{ 17}$,
A~Woolverton$^{ 13}$, 
M~Yeh$^{ 1}$,
C~Zhang$^{ 1}$ and
X~Zhang$^{ 15}$ \\
(The PROSPECT Collaboration)
}

\address{$^{1}$Brookhaven National Laboratory, Upton, NY, USA} \vspace{-0.4\baselineskip}
\address{$^{2}$Department of Physics and Astronomy, University of Hawaii, Honolulu, HI, USA} \vspace{-0.4\baselineskip}
\address{$^{3}$Department of Physics and Astronomy, University of Tennessee, Knoxville, TN, USA} \vspace{-0.4\baselineskip}
\address{$^{4}$Department of Physics, Boston University, Boston, MA, USA} \vspace{-0.4\baselineskip}
\address{$^{5}$Department of Physics, Drexel University, Philadelphia, PA, USA} \vspace{-0.4\baselineskip}
\address{$^{6}$Department of Physics, Illinois Institute of Technology, Chicago, IL, US} \vspace{-0.4\baselineskip}
\address{$^{7}$Department of Physics, Le Moyne College, Syracuse, NY, USA} \vspace{-0.4\baselineskip}
\address{$^{8}$Department of Physics, Temple University, Philadelphia, PA, USA} \vspace{-0.4\baselineskip}
\address{$^{9}$Department of Physics, University of Wisconsin, Madison, WI, USA} \vspace{-0.4\baselineskip}
\address{$^{10}$Department of Physics, United States Naval Academy, Annapolis, MD, USA} \vspace{-0.4\baselineskip}
\address{$^{11}$George W.\,Woodruff School of Mechanical Engineering, Georgia Institute of Technology, Atlanta, GA, USA} \vspace{-0.4\baselineskip}
\address{$^{12}$High Flux Isotope Reactor, Oak Ridge National Laboratory, Oak Ridge, TN, USA} \vspace{-0.4\baselineskip}
\address{$^{13}$Institute for Quantum Computing and Department of Physics, University of Waterloo, Waterloo, ON, Canada} \vspace{-0.4\baselineskip}
\address{$^{14}$National Institute of Standards and Technology, Gaithersburg, MD, USA} \vspace{-0.4\baselineskip}
\address{$^{15}$Nuclear and Chemical Sciences Division, Lawrence Livermore National Laboratory, Livermore, CA, USA} \vspace{-0.4\baselineskip}
\address{$^{16}$Physics Division, Oak Ridge National Laboratory, Oak Ridge, TN, USA} \vspace{-0.4\baselineskip}
\address{$^{17}$Wright Laboratory, Department of Physics, Yale University, New Haven, CT, USA} \vspace{-0.4\baselineskip}

\ead{prospect.collaboration@gmail.com}

%% file: abstract.tex
The Precision Reactor Oscillation and Spectrum Experiment, PROSPECT, has made world-leading measurements of reactor antineutrinos at short baselines. In its first phase, conducted at the High Flux Isotope Reactor (HFIR) at Oak Ridge National Laboratory, PROSPECT produced some of the strongest limits on eV-scale sterile neutrinos, made a precision measurement of the reactor antineutrino spectrum from $^{235}$U, and demonstrated the observation of reactor antineutrinos in an aboveground detector with good energy resolution and well-controlled backgrounds. The PROSPECT collaboration is now preparing an upgraded detector, PROSPECT-II, to probe yet unexplored parameter space for sterile neutrinos and contribute to a full resolution of the Reactor Antineutrino Anomaly, a longstanding puzzle in neutrino physics. By pressing forward on the world's most precise measurement of the $^{235}$U antineutrino spectrum and measuring the absolute flux of antineutrinos from $^{235}$U, PROSPECT-II will sharpen a tool with potential value for basic neutrino science, nuclear data validation, and nuclear security applications. Following a two-year deployment at HFIR, an additional PROSPECT-II deployment at a low enriched uranium reactor could make complementary measurements of the neutrino yield from other fission isotopes. PROSPECT-II provides a unique opportunity to continue the study of reactor antineutrinos at short baselines, taking advantage of demonstrated elements of the original PROSPECT design and close access to a highly enriched uranium reactor core.

%% file: Introduction.tex
\section{Introduction}

The first phase of the Precision Reactor Oscillation and Spectrum Experiment, PROSPECT, produced the world's most precise measurement of the antineutrino energy spectrum from a highly enriched uranium (HEU) reactor and set new limits on active-sterile neutrino mixing in the eV-scale range of mass splitting \cite{prospect_osc, prospect_spec, prospect_prd}. The PROSPECT collaboration is now preparing an upgraded detector, PROSPECT-II, for a second phase of measurements. As in the original PROSPECT design, the PROSPECT-II detector will contain a segmented $^6$Li-doped liquid scintillator volume optimized for inverse beta decay detection with minimal cosmic-ray shielding. The PROSPECT-II detector design introduces evolutionary changes that improve detector stability and longevity. PROSPECT-II anticipates a two-year run at the High Flux Isotope Reactor (HFIR) at Oak Ridge National Laboratory, in a location covering baselines (distances from reactor core to detector volume) of 7-9 m. 
The PROSPECT-II deployment at HFIR will
expand the statistical power of the original PROSPECT dataset by more than an order of magnitude. An additional deployment of the same detector at a low enriched uranium (LEU) reactor would allow a systematics-correlated measurement of antineutrinos from other fissioning isotopes.  

Data from PROSPECT-II will address multiple goals of interest to the particle physics, nuclear science, and nuclear security communities:

\begin{itemize}
    \item \textbf{Particle physics impacts:} \uline{PROSPECT-II will search unexplored areas of phase space for possible active-sterile neutrino mixing}.   While searching in the $\sim$1-20~eV$^2$ mass splitting region, the experiment will provide one important component of a resolution to the Reactor Antineutrino Anomaly~\cite{bib:mueller2011,bib:mention2011}, one of the short-baseline anomalies in neutrino physics \cite{AnomalyWhite}. Beyond the Reactor Antineutrino Anomaly, PROSPECT-II will continue to provide unique sensitivity to the electron-flavor disappearance oscillation channel, complementary to searches for new physics with muon-flavor sources.  In particular, PROSPECT-II will have unique sensitivity to a region of phase space that has been raised as a possible complication for interpretation of long-baseline CP violation measurements \cite{Kayser, cp_kelly}.
    
    \uline{PROSPECT-II will measure the absolute flux and spectrum of antineutrinos produced by an HEU reactor and potentially an LEU reactor}. These measurements will advance knowledge of reactors as a general tool for neutrino physics, with potential value for precision studies of the neutrino mass hierarchy~\cite{juno2} and searches for beyond-Standard Model (BSM) interactions using coherent elastic neutrino-nucleus scattering (CEvNS)~\cite{Ricochet,Connie,conus}. 
    An absolute flux measurement at an LEU reactor would also add a data point to the set of measurements testing whether the data-model discrepancy of the Reactor Antineutrino Anomaly varies by fuel isotope~\cite{bib:prl_evol, reno_evol}.

    \item \textbf{Nuclear science impacts:}  \uline{PROSPECT-II will push the precision on direct knowledge of the $^{235}$U antineutrino spectrum below that of currently claimed model uncertainties}, offering a view on the set of fission product beta decays that give rise to this spectrum.  
    When combined with existing LEU data, PROSPECT-II's measurement can help clarify the nature of discrepancies between measured and predicted reactor spectra, which may derive from inaccuracies in nuclear databases or shortcomings in the current understanding of forbidden beta decays~\cite{sonzogni_insights, vogel_review}.
    By following a HFIR deployment with measurements at a LEU reactor, \uline{PROSPECT-II can deliver systematics-correlated HEU and LEU flux and spectrum measurements}, which will sharpen global understanding of neutrino emission from $^{239}$Pu fissions and from the minor fission contributors $^{238}$U and $^{241}$Pu.
   By enabling high-level, aggregate tests of fission yield and beta decay databases, neutrino flux and spectrum measurements from experiments such as PROSPECT-II offer a new mode of validation in the nuclear data pipeline.

    \item \textbf{Nuclear security impacts:} \uline{PROSPECT-II will advance both neutrino physics knowledge and detection capabilities to enable applications in nuclear security.}  Possible applications include cooperative verification of a reactor shutdown agreement or safeguards for advanced reactor designs~\cite{Bernstein:2019hix}. While such concepts are still early in development, it is clear that any use of neutrino detectors in reactor safeguards or verification regimes would benefit from the precise characterization of reactor neutrino source terms that PROSPECT-II will produce. The first run of PROSPECT has demonstrated the capability to suppress backgrounds with minimal overburden ($<1$ m water equivalent). Upgrades in detector robustness compared to the original PROSPECT design, and portability to on-surface locations at multiple reactor sites, will demonstrate additional applications-relevant features.
\end{itemize}

The following section describes the physics context motivating data collection with the PROSPECT-II detector.  
This background material includes a description of the theoretical possibility of active-sterile neutrino mixing, a survey of the current status of efforts to explain data-model discrepancies in reactor neutrino experiments, and a discussion of recently highlighted connections between long-baseline oscillation experiments, short-baseline oscillation experiments, and broader attempts to probe Standard Model and BSM physics.  


\section{Physics context for PROSPECT-II}
\label{sec:Introduction}

\subsection{The possibility of active-sterile neutrino mixing}
\label{subsec:sterile_big}

Since the late 1990s, observations of neutrino oscillations have established the presence of flavor mixing in the neutrino sector and shown that neutrinos ($\nu$) have mass.  
The existence of nonzero neutrino mass suggests the existence of one or more right-handed neutrinos which do not participate in weak interactions, and these `sterile' neutrino states $\nu_s$ are indeed a common facet of Standard Model extensions incorporating non-zero neutrino masses~\cite{AnomalyWhite}.  
Neutrino mass generation models generally do not show a preference toward specific mass ranges or active-sterile couplings for $\nu_s$~\cite{ deGouvea:2009fp}.  
As a result, an array of techniques, sensitive to a vast range of mass scales, have been employed to search for signatures of their existence.  
Above roughly 1~MeV, experimental limits on active-sterile couplings are dominated by searches for production or decay of sterile particles, while below 1~MeV, limits are dominated by either indirect astrophysical/cosmological observations or direct searches for signatures of active-sterile neutrino flavor mixing~\cite{Bolton:2019pcu}.  

At low mass scales, sterile neutrinos can create observable effects by mixing with active neutrino and antineutrino flavors, $\nu_e$ (\nuebar), $\nu_{\mu}$ ($\overline{\nu}_{\mu}$), and $\nu_{\tau}$ ($\overline{\nu}_{\tau}$) in a manner analogous to the mixing of the three active flavors with one another.  
In a minimal model including one sterile state that mixes with the three active flavor states (a `3+1' model), the $3\times3$ Pontecorvo-Maki-Nakagawa-Sakata (PMNS) neutrino flavor mixing matrix expands to a $4\times4$ matrix~\cite{giunti_review}:

\begin{eqnarray}
\label{eq:PMNSMatrix}
U_{PMNS} = \begin{pmatrix} U_{e1} & U_{e2} & U_{e3} & U_{e4} \\ U_{\mu1} & U_{\mu2} & U_{\mu3} & U_{\mu4} \\ U_{\tau1} & U_{\tau2} & U_{\tau3} & U_{\tau4} \\ U_{s1} & U_{s2} & U_{s3} & U_{s4} \\ \end{pmatrix},
\end{eqnarray}
\vspace{1mm} 

where the fourth row and column govern active-sterile mixing.  
The matrix elements in this row and column are combinations of three new mixing angles,  $\theta_{14}$, $\theta_{24}$, and $\theta_{34}$, as well as two new CP violating phases, $\delta_{14}$ and $\delta_{34}$; for example, sin$^2$2$\theta_{14}$ = 4$|$U$_{e4}|^2$(1-$|$U$_{e4}|^2$).  

The expanded mixing matrix may give rise to observable oscillation effects  beyond those established between the three active flavors~\cite{pdg_2018}.  
For example, a source of \nuebar could show an additional baseline- and energy-dependent disappearance pattern beyond the known patterns driven by $\theta_{12}$~\cite{KamLAND_shape} and $\theta_{13}$~\cite{bib:prl_shape,bib:reno_shape,dc_nature}. In a minimal 3+1 model including an eV-scale sterile state, sterile-driven, short-baseline \nuebar disappearance would be characterized by the survival probability
\begin{equation}
P_{\nuebar \rightarrow \nuebar} \approx 1 - \sin^22\theta_{14} \sin^2 \frac{\Delta m^2_{41} L}{4E},   
\label{eq:survprob}
\end{equation}
where $\theta_{14}$ is the new mixing angle, $\Delta m^2_{41}$ is the mass splitting between the new fourth state and the lightest of the active neutrino states (assuming the normal mass hierarchy), $L$ is the distance between antineutrino source and detector (called the experiment baseline), and $E$ is the antineutrino energy.  

A challenge for experiments seeking to observe active-sterile oscillations is that, as mentioned above, there is little \textit{a priori} constraint on the values of possible active-sterile mixing angles and mass splittings.    
While neutrino mass generation mechanisms in super-symmetric or grand unification theories suggest that sterile states could have masses as large as the electro-weak~\cite{Casas:2004gh} or GUT scales~\cite{Davidson:2008bu}, other mechanisms such as low-energy seesaws~\cite{Donini:2011jh,Fan:2012ca,Blennow:2011vn} 
describe sterile neutrino states in the eV to GeV mass regime.  
Without knowing the mass of the sterile states, there is no theoretical indication of the wavelength of active-sterile oscillation effects of Equation \ref{eq:survprob}, and thus no indication of which $L$ and $E$ to experimentally target.

\subsection{Short-baseline anomalies and eV-scale neutrino searches}
\label{subsec:sterile_ev}

In the past few decades, a series of anomalies observed in short-baseline neutrino experiments have collectively directed a spotlight on a particular region of sterile neutrino phase space~\cite{AnomalyWhite}.  
The accelerator experiments LSND~\cite{lsnd} and MiniBooNE~\cite{mboone} saw excesses of $\nu_e$ ($\overline{\nu}_e$)-like events in $\nu_{\mu}$ ($\overline{\nu}_{\mu}$) beams.  
The anomalous presence of electron-like events in these experiments could be explained by $\nu_s$-driven $\mu$-to-$e$ flavor oscillations with an active-sterile mass splitting on the order of 1 eV$^2$.  
The establishment of the eV-scale mass regime as an area of interest was reinforced by the observation of anomalously low $\nu_e$ interaction rates in solar neutrino experiments observing high-intensity radioactive electron capture decay sources~\cite{gallium}; 
these results could also be explained by eV-scale $\nu_s$. In 2011, another anomaly appeared in connection with reactor \nuebar experiments~\cite{bib:mueller2011,bib:mention2011}. This effect, known as the Reactor Antineutrino Anomaly, is explained in greater detail in the following section. It too could be explained by oscillations involving a $\nu_s$ with an eV-scale mass.   

Together, these observations spurred dozens of new proposed experimental efforts worldwide~\cite{AnomalyWhite}, with more than a dozen eventually taking neutrino physics data~\cite{giunti_review}. 
Their number reflects the monumental impact that a sterile neutrino discovery would have on particle physics, as a major step beyond the Standard Model and possibly a clue to other BSM phenomena.  

\subsection{The Reactor Antineutrino Anomaly}

A primary motivation for PROSPECT was a discrepancy between predicted and observed reactor antineutrino interaction rates known as the Reactor Antineutrino Anomaly.  
The observations date back to a series of ton-scale experiments in the 1980s and 1990s, each measuring the inverse beta decay (IBD) interaction rate, $\nuebar + p \rightarrow e^+ + n$, of \nuebar emitted by a commercial or research reactor~\cite{bib:ILL_nu, bib:gosgen, bib:Krasno1, bib:rovnoCC, bib:Rovno2, bib:B4, Declais:1994su, Krasnoyarsk2:1994ut, bib:srp}.  
These observations, made at reactor-detector baselines ranging from roughly 10 to 100 meters, were consistent with \nuebar predictions of their day, which were derived from $\beta^-$ spectrum measurements of fissioning isotopes \cite{bib:ILL_1, bib:ILL_2}.  

\begin{figure}[hptb!]
\centering
\includegraphics[trim = 0.0cm 0.0cm 0.0cm 0.0cm, clip=true, width=\textwidth]{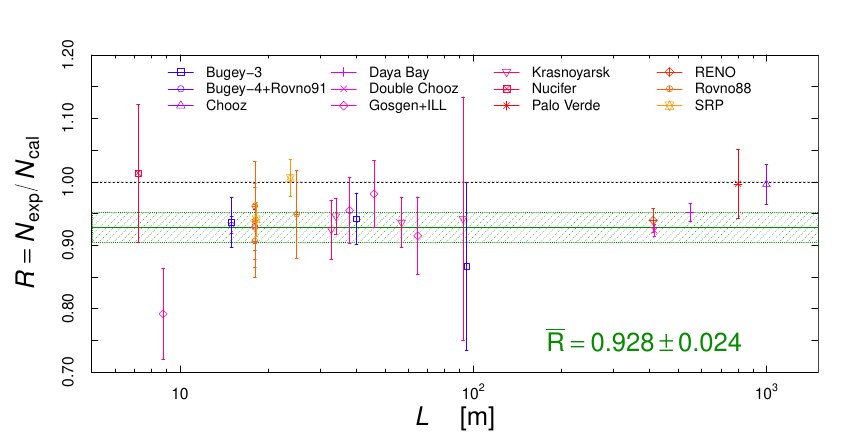}
\caption{Measurements of reactor \nuebar flux from a variety of neutrino experiments, provided relative to conversion-predicted flux models~\cite{bib:huber,bib:mueller2011}.  The offset between the predicted (black) and globally-averaged measured (green) flux is referred to as the Reactor Antineutrino Anomaly. The green band includes uncertainties from the experimental measurements (accounting for correlations) and the theoretical uncertainty quoted on the conversion-predicted flux models~\cite{bib:huber,bib:mueller2011}.  Figure from Ref.~\cite{giunti_review}.  
}
\label{fig:raa}
\end{figure}

The Reactor Antineutrino Anomaly appeared when the predicted flux of reactor \nuebar was recalculated in 2011 using the same $\beta^-$ spectrum-based, or `conversion,' approach. The updated calculations used the same $\beta^-$ spectra as previous calculations but included updated nuclear information and an updated estimate of the neutron lifetime. 
Two independent efforts put the newly-calculated observable neutrino flux roughly 6\% higher than previous estimates \cite{bib:mueller2011, bib:huber}.  
A representation of \nuebar flux measurements relative to this new estimate is shown in Figure~\ref{fig:raa}.  The shift in predicted flux opened a gap between observations and predictions, making them incompatible with a significance at 98.6\% confidence level \cite{bib:mueller2011}.  
The existence of this gap was subsequently confirmed at longer ($>$100~m) baselines in a new generation of reactor-based $\theta_{13}$ experiments~\cite{bib:prl_reactor,reno_evol,dc_nature}.  

\subsection{Sterile neutrinos as the source of the Reactor Antineutrino Anomaly}
\label{subsec:raa_steriles}

 One way to explain the data-model gap of the Reactor Antineutrino Anomaly is to introduce mixing between the three Standard Model neutrinos and $\nu_s$ states.  
In the minimal 3+1 scenario described in previous sections, \nuebar disappearance is governed by the expanded PMNS mixing matrix parameter $U_{e4}$ (\emph{i.e.}~$\theta_{14}$) and the active-sterile neutrino mass difference  $\Delta m^2_{41}$, with \nuebar disappearance as a function of baseline described by Equation~\ref{eq:survprob}.  
The Reactor Antineutrino Anomaly could be explained by such disappearance with an amplitude of $\sin^22\theta_{14} \sim 10^{-1}$, twice the observed data-model gap, and a mass splitting of $\Delta m^2_{41}~\sim$~1 eV$^2$ or greater, large enough to average oscillations into an energy-independent deficit at baselines above 10~m.  
The comparatively larger deficit in the shortest-baseline data-point from the ILL experiment~\cite{bib:ILL_nu,ILL_fix} drives a mildly favored best-fit point to the lower range of these mass splittings, while the majority of the suggested phase space appears at higher mass splittings.  

Intriguingly, the mixing angle and mass splitting range indicated by the Reactor Antineutrino Anomaly roughly match parameter ranges indicated by the previously described anomalies in radioactive source and accelerator neutrino experiments.  
Thus, when it appeared, the Reactor Antineutrino Anomaly became a significant link in a chain of empirical motivators for eV-scale $\nu_s$ searches.   

In its first phase, the PROSPECT experiment pursued the sterile neutrino explanation for the Reactor Antineutrino Anomaly by searching directly for the oscillatory behavior described in Equation \ref{eq:survprob}.  
At reactor-based experiments, the $L/E$ character of an oscillation-produced \nuebar deficit appears as a baseline-dependent 
variation in the detected $\nuebar$ spectrum.
By designing a system to measure IBD interactions over a range of baselines, PROSPECT could search for such effects.
At just under 10~m from the HFIR reactor core, PROSPECT was most sensitive to oscillations corresponding to neutrino mass splittings of roughly 1-10 eV$^2$, overlapping well with the Reactor Antineutrino Anomaly best fit region.  

\begin{table*}[thbp!]
\centering
\begin{tabular}{c||c|c|c}
\hline
Experiment & Reactor Type & Scintillator Technology & Search Strategy \\ \hline \hline 
Bugey-3~\cite{Declais:1994su} & Power LEU & Segmented $^6$Li PSD Liquid & Multi-Site \\ \hline 
NEOS~\cite{bib:neos}  & Power LEU & Single-Volume Gd PSD Liquid & Single-Site \\ \hline
DANSS~\cite{danss_osc}  & Power LEU & Segmented Plastic & Multi-Site \\ \hline
STEREO~\cite{stereo_2018}  & Compact HEU & Segmented Gd PSD Liquid & Multi-Zone \\ \hline
PROSPECT~\cite{prospect_osc}  & Compact HEU & Segmented $^6$Li PSD Liquid & Multi-Zone \\ \hline
Neutrino-4~\cite{bib:neutrino4_osc}  & Compact HEU & Segmented Gd Liquid & Multi-Site/Zone \\ \hline

\end{tabular}
\caption{Comparison of experimental parameters of performed reactor-based sterile neutrino search experiments. `Multi-site' means deployment of a detector in multiple locations relative to the reactor. Experiments are listed in rough order of increasing first oscillation publication date.  See text for further details.  
}
\label{tab:current_osc}
\end{table*}

A number of other experiments worldwide have also pursued spectrum-based short-baseline reactor \nuebar oscillation searches using a variety of reactor types, scintillator technologies, and experimental strategies.  
A summary of the experimental parameters of these searches appears in Table~\ref{tab:current_osc}.  HFIR's compact core enables PROSPECT to achieve superior coverage at high $\Delta m^2_{41}$ relative to other reactor experiments, as shorter-wavelength high-$\Delta m^2_{41}$ oscillation features are less washed out by the range in sampled baselines produced by the finite core geometry~\cite{VSBL}.  
The use of $^6$Li-doped, PSD-capable liquid scintillator and detector segmentation enables PROSPECT to achieve a level of active background rejection so far unmatched by other technology choices, allowing a sensitive overburden-free oscillation search.  
Finally, PROSPECT's multi-zone detector enables simultaneous sampling of \nuebar at many energies and baselines in a single detector deployment, without major systematics from the detector's absolute energy response and the absolute rate and energies of \nuebar produced by HFIR.  

\begin{figure}[hptb!]
\centering
\includegraphics[trim = 0.0cm 0.0cm 0.5cm 0.0cm, clip=true, width=0.64\textwidth]{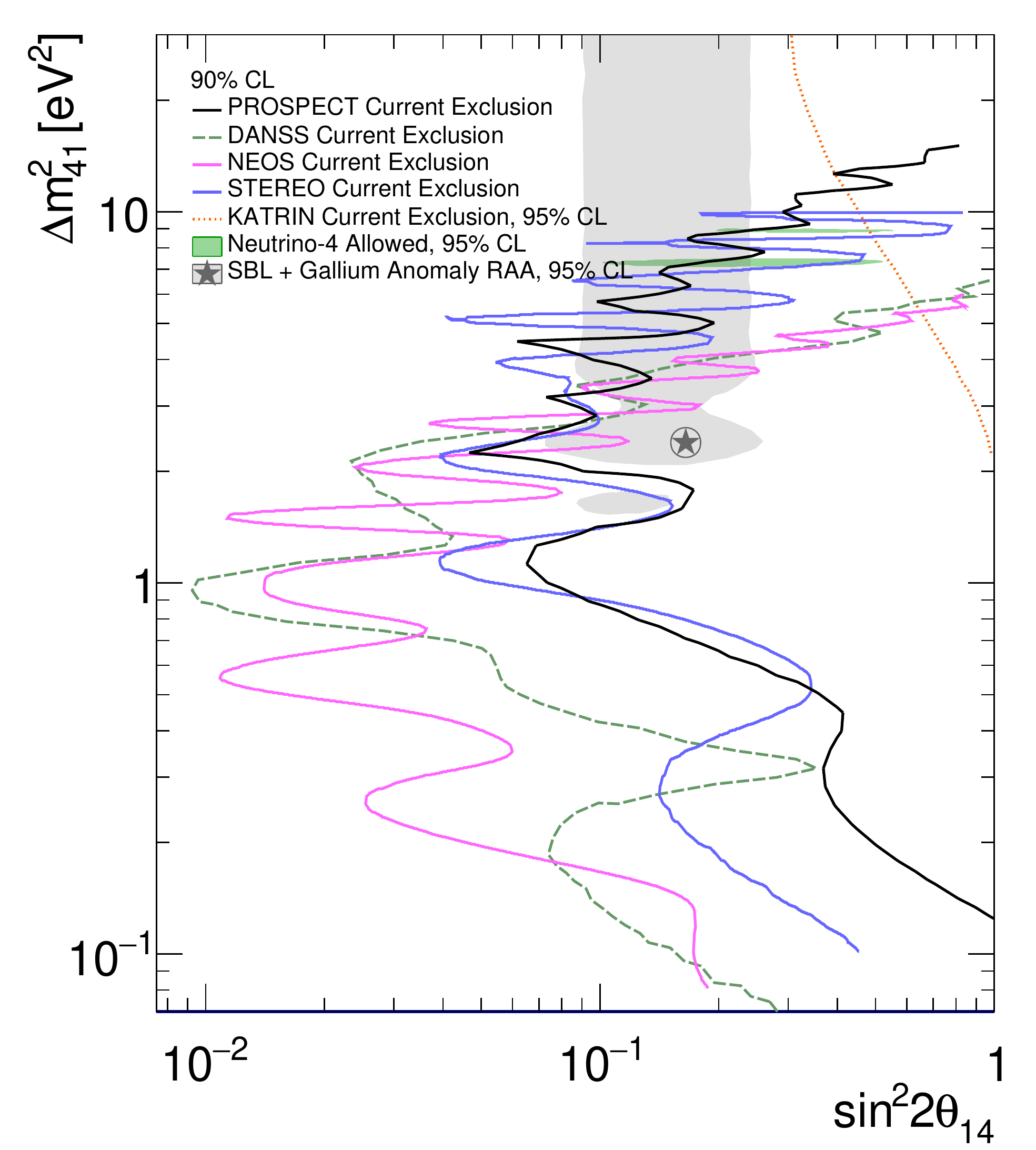}
\caption{Sterile neutrino parameter space exclusion regions obtained by recent short-baseline reactor neutrino~\cite{danss_osc,bib:neos,stereo_2019,prospect_prd} and tritium beta decay~\cite{katrin_sterile} experiments.  Also shown are the parameter space allowed by the sterile neutrino hypothesis for the Reactor Antineutrino Anomaly, according to~\cite{giunti_diagnose}, and the 95\% allowed region claimed in controversial Neutrino-4 results~\cite{bib:neutrino4_osc}.  While phase space has been ruled out at low mass splitting, unaddressed regions remain above $\sim$5~eV$^2$.}
\label{fig:current_osc}
\end{figure}

To date, nearly all of the short-baseline reactor experiments see no substantial evidence of oscillations.  
As Figure~\ref{fig:current_osc} illustrates, the null results of recent oscillation searches in PROSPECT \cite{prospect_prd}, NEOS \cite{bib:neos}, DANSS \cite{danss_osc}, and STEREO \cite{stereo_2019} were sufficient to exclude the best-fit sterile neutrino parameters for the Reactor Antineutrino Anomaly, along with a sizable portion of the surrounding eV-scale phase space.  
One short-baseline neutrino experiment, in contrast, has claimed to observe oscillation signals. The Neutrino-4 experiment reports a signal corresponding to sin$^2$2$\theta_{14}$ = 0.36 and $\Delta m^2_{41}$ = 7.3 eV$^2$ using controversial analysis methods~\cite{bib:neutrino4_osc}.  
While concerns have been raised regarding the Neutrino-4 data and analysis~\cite{n4_comment,giunti_mc,Coloma:2020ajw}, the phase space region in question has not yet been directly addressed with data from other experiments.  
This is because the region lies in a higher-$\Delta\textrm{m}^2$ portion of oscillation phase space on the margins of current STEREO and PROSPECT sensitivity and below the accessible range of current tritium $\beta^-$ decay endpoint experiments, such as KATRIN~\cite{katrin_sterile}.  

In summary, short-baseline reactor experiments have excluded a large portion of the oscillation phase space that could explain the Reactor Antineutrino Anomaly, but current results leave open a possible solution space at higher $\Delta m^2_{41}$, above roughly 5~eV$^2$. 
Future reactor-based experiments, in combination with future KATRIN data, can address this remaining region of phase space to provide complete closure on the sterile neutrino interpretation of the Reactor Antineutrino Anomaly. Section \ref{subsec:phys_osc} presents the sensitivity of PROSPECT-II in this region.

\subsection{Flux predictions as the source of the Reactor Antineutrino Anomaly}
\label{subsec:raa_flux}

An alternative to the sterile neutrino hypothesis for the Reactor Antineutrino Anomaly is inaccuracy in the prediction of the reactor neutrino flux. In general, the production of neutrinos in a reactor per unit neutrino energy $E$ and time $t$ is modeled as:
\begin{equation}
\frac{dN(E,t)}{dE dt} = \frac{P_{th}(t)}{\overline{E_f}(t)} \sum_{i=1}^4 f_i(t) s_i(E)c_i^{ne}(E)
\label{eq:flux}
\end{equation}
where $P_{th}$ is the reactor thermal power, $f_i$ is the fraction of fissions occurring on isotope $i$, $s_i$ is the neutrino flux per fission from isotope $i$, $c_i^{ne}$ denotes non-equilibrium contributions from long-lived fission fragments, and $\overline{E_f} = \sum_i f_i(t)e_i$ is the average energy released per fission, with $e_i$ denoting the energy released by isotope $i$. The index $i$ runs over the major contributions to fission in a given reactor type. In a reactor fueled with low enriched uranium (LEU), including most commercial power reactors, these isotopes are $^{235}$U, $^{239}$Pu, $^{238}$U, and $^{241}$Pu. PROSPECT has observed the HFIR core, which burns highly enriched uranium (HEU) fuel and generates a neutrino flux $\sim99\%$ from $^{235}$U fissions. Antineutrino emission from non-fuel isotopes \cite{nonfuel} and nearby spent fuel \cite{bib:fengpeng, bib:zhoubin} may also contribute to the signal observed in reactor-based experiments. In PROSPECT, non-fuel isotopes contribution about 1\% of the antineutrino signal, and spent fuel contributions are negligible. 

The Reactor Antineutrino Anomaly could be explained by a systematic bias in the absolute normalization of one or more parameters in Equation \ref{eq:flux}. Reactor power, fission fractions, and per-fission energy are relevant to reactor operation and have been well established through multiple mature measurement techniques. Systematics related to these quantities result in neutrino flux uncertainties estimated at a few percent or less~\cite{bib:cpc_reactor}, making them unlikely to contribute significantly to the observed anomaly.  
This leaves the $s_i$, the neutrino flux per fission of each isotope, as a potential source of bias affecting the global reactor neutrino picture.  
Since conversion-based predictions of $s_i$ depend on some shared and some unique inputs across fission isotopes, a systematic effect may be common to all isotopes or specific to a subset of them.  
As an explanation for the Reactor Antineutrino Anomaly, the flux prediction hypothesis is distinct from the sterile neutrino hypothesis in that it does not generate $L/E$-dependent effects.  
It is also distinct in that the data-prediction gap it creates may depend on the fission fractions in a reactor, which in LEU reactors change over time. 

\begin{figure}[hptb!]
\centering
\includegraphics[trim = 0.0cm 7cm 37.5cm 0.0cm, clip=true, width=0.7\textwidth]{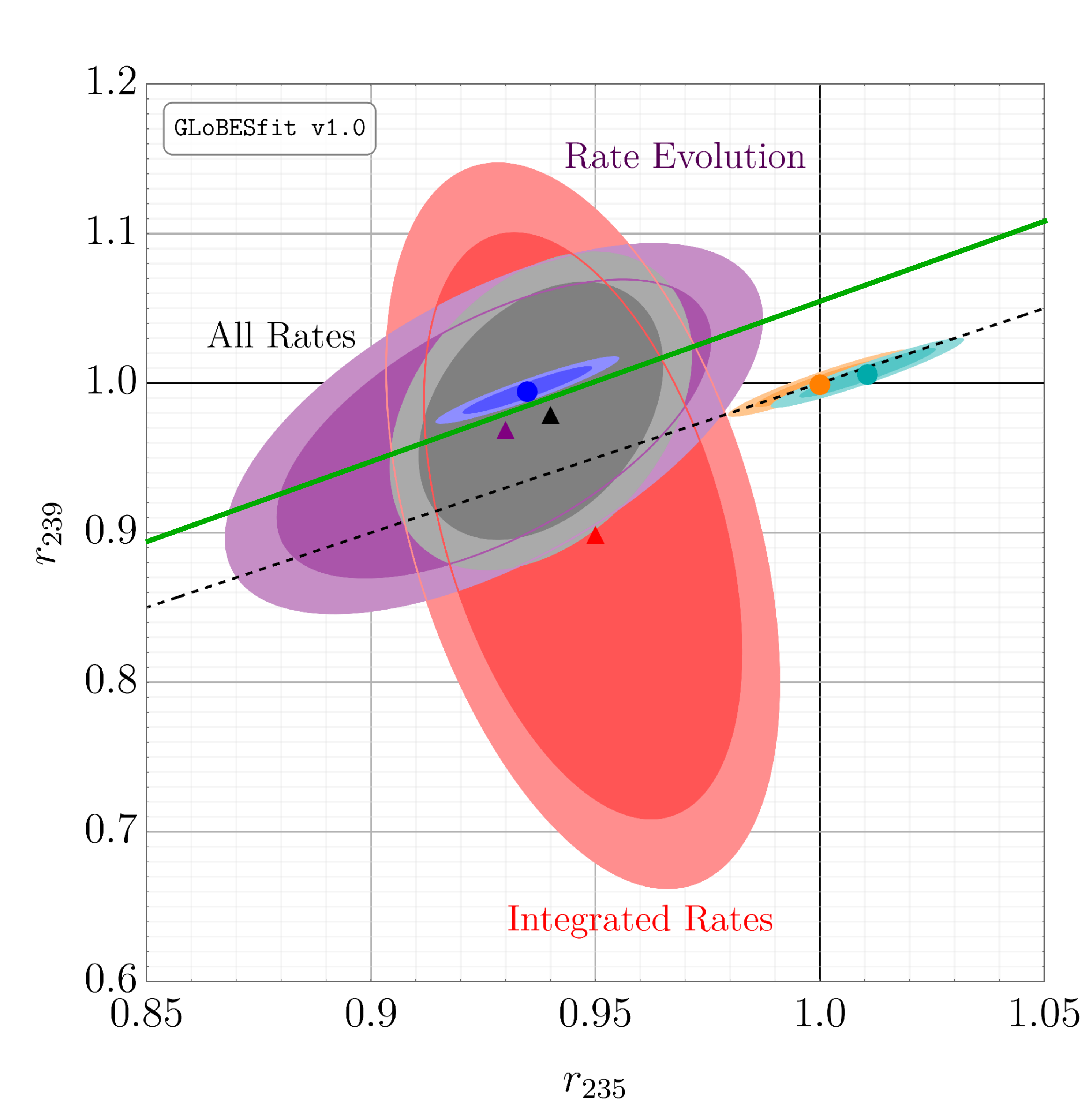}
\caption{95\% and 99\% confidence-level allowed regions for \uFive~ and \pNine~ IBD yields derived from Daya Bay and RENO flux evolution measurements (purple), other global reactor flux datasets (red), and their combination (gray), as well as 1$\sigma$ and 2$\sigma$ allowed regions from two conversion-based reactor flux models (orange~\cite{bib:huber} and cyan~\cite{hayen_initio}) and one \emph{ab initio} model (blue~\cite{bib:fallot2}, drawn with the conversion model uncertainties).  The black dashed line is the $r_{235}=r_{239}$ diagonal fixed to the prediction in~\cite{bib:huber}. The green solid line corresponds to the recent \uFive-\pNine~fission beta ratio measurement from Ref.~\cite{kopeikin2021}.  All yields are given as a ratio with respect to the conversion predictions of Ref.~\cite{bib:huber}.  Figure altered from Ref.~\cite{huber_berryman}.}
\label{fig:intro_flux}
\end{figure}

In the last five years, the flux prediction hypothesis has been bolstered by experimental and theoretical work from both the nuclear and neutrino physics communities. 
In the Daya Bay experiment, analyzing the evolution of neutrino rate with fuel composition shows a data-prediction gap that decreases with decreasing $^{235}$U fission fraction \cite{bib:prl_evol}.  
The RENO experiment has produced flux evolution results consistent with those of Daya Bay~\cite{reno_evol}.  
As shown in Figure~\ref{fig:intro_flux}, these flux evolution results can be mapped into isotopic IBD yield, $\sigma_i = \int s_i(\textrm{E}) \sigma_{IBD}(\textrm{E}) d$E, for the individual isotopes $i$ = $^{235}$U and $^{239}$Pu, where E is neutrino energy and $\sigma_{IBD}$ is the IBD cross-section.  
Relative to the canonical conversion predictions (orange region), flux evolution results (purple region) prefer a larger IBD yield deficit for \uFive~than for \pNine.  
As noted above, these isotope-dependent measured neutrino rate deficits are unique to the flux prediction hypothesis. Put another way, if the sterile neutrino hypothesis explains the Reactor Antineutrino Anomaly, and if the canonical conversion predictions (orange point) are accurate, observations should fall on the dashed black line in Figure~\ref{fig:intro_flux} instead of above it. 

On the nuclear experiment side, a recent measurement of the post-fission beta decay ratio in $^{235}$U vs. $^{239}$Pu, illustrated by a green line in Figure~\ref{fig:intro_flux}, lends further support to the flux prediction hypothesis~\cite{kopeikin2021}.  
This measurement indicates that the 30-year-old $\beta^-$ yield measurements used to predict the reactor neutrino spectra are inaccurate in a way that leads to a $\sim$5\% overestimate of the $^{235}$U neutrino yield relative to $^{239}$Pu.  
Possible biases in the original $\beta^-$ yield measurements have been independently noted by others~\cite{bib:ILL_calib}.   
While the systematic uncertainties associated with the new beta ratio study are not quantified by the authors, the central value is consistent with the Daya Bay measurement of $^{235}$U and $^{239}$Pu yields and HEU-based measurements of the $^{235}$U yield.  
Notably, unlike conversion predictions, \textit{ab initio} predictions of $s_i$ made using standard fission yield and nuclear structure databases (blue region in Figure~\ref{fig:intro_flux}) generate IBD yields that are generally consistent with the Daya Bay flux evolution results~\cite{hayes_evol} and continue to be so even as better $\beta^-$ feeding measurements arrive from a variety of nuclear physics experiments~\cite{bib:fallot2,tas_few,tas_lots,tas_rbi,tas_rbbr,tas_br,tas_nb,tas_rb}.  

In sum, experimental and theoretical developments in the past five years point to a flux prediction issue in canonical conversion models.  
Still, knowledge of the reactor neutrino flux remains blurry, and observations do not yet guarantee that oscillation plays a minor role in discrepancies between predicted and measured fluxes.  
The 
possible space for 
oscillations is illustrated in studies of `hybrid' models including both flux prediction issues and oscillations~\cite{giunti_evol, giunti_diagnose}. 

Because of the longstanding role of the Reactor Antineutrino Anomaly in motivating beyond-Standard Model theory and experimental searches in the neutrino sector, the particle physics community has a major stake in securing its definitive resolution, whatever the underlying cause.  
An authoritative resolution should be possible through the collection of additional, complementary information from both absolute reactor flux measurements and oscillation searches.  
The second phase of PROSPECT, outlined in Section \ref{sec:PII_PhysicsGoals}, will make important contributions to this resolution. 

\subsection{A broader view of sterile neutrino searches with reactors}
\label{subsec:sterile_broad}

As described above, the phase space available for sterile neutrinos to explain the Reactor Antineutrino Anomaly has narrowed considerably since PROSPECT started taking data.
However, as introduced in Sections~\ref{subsec:sterile_big} and~\ref{subsec:sterile_ev}, the subject of sterile neutrinos is broader than the question of what explains that particular anomaly.  
Even if a flux prediction issue is responsible for the data-observation gap in the reactor sector, short-baseline anomalies in the accelerator sector may be explained by BSM physics.  
Some of these new physics models could be further explored with data from PROSPECT-II.  
Additionally, it has been emphasized that active-sterile neutrino mixing, whether related to short-baseline anomalies or not, could complicate the interpretation of long-baseline CP violation experiments. As explained below, PROSPECT-II has unique sensitivity to help address this possibility.  

Returning to the expanded PMNS matrix presented in Equation~\ref{eq:PMNSMatrix}, and focusing on the decades of mass-splitting range from $\Delta$m$^2_{32}$ to tens of eV$^2$: 
\begin{itemize}
\item $U_{e4}$ limits are dominated by short baseline (few hundred meters or less) reactor measurements of \nuebar disappearance. In this context, improvements in $\theta_{14}$ coverage from PROSPECT-II and the other short-baseline reactor experiments DANSS~\cite{danss_osc}, NEOS-II~\cite{neos2}, STEREO~\cite{stereo_2019}, and TAO~\cite{juno_tao} will help drive the field on the 5-10 year timescale.  
\item $U_{\mu4}$ limits are primarily from $\nu_{\mu}$ disappearance measurements at meson decay-in-flight accelerator neutrino beam experiments, such as MINOS+~\cite{minos}, and from measurements of atmospheric neutrinos, as performed by IceCube~\cite{IceCube_sterile}.  
\item Short-baseline accelerator electron-like appearance searches, like those at Fermilab's SBN Program~\cite{sbn}, probe multiple matrix elements and effects related to their combination, such as CP violation~\cite{sbn_cpv}.  
\end{itemize}
In the previously described minimal 3+1 sterile oscillation case, matrix elements $U_{e4}$ and $U_{\mu4}$ are defined by the mixing angles $\theta_{14}$ and $\theta_{24}$. Anomalous appearance or disappearance results may just as easily reflect additional physics processes beyond the minimal 3+1 oscillations described by Equation~\ref{eq:PMNSMatrix}, such as non-standard neutrino interactions~\cite{nsi},  neutrino decay~\cite{PalomaresRuiz:2005vf,decay_conrad,decay_deg,decay_machado}, or some combination of processes.  
The existence of multiple sterile neutrino states would expand the dimensions of Equation~\ref{eq:PMNSMatrix} and introduce additional flavor mixing and CP-violating parameters~\cite{bib:kopp, vsbl_mult}.  
If right-handed neutral leptons are required to explain non-zero neutrino mass, it may indeed seem natural to have multiple sterile neutrino states.

To see how PROSPECT-II can inform interpretations of anomalies from other interaction channels, it is useful to first consider how minimal and `non-minimal' 3+1 scenarios impact short-baseline datasets.  
In the minimal 3+1 case, the $\nu_{\mu} \rightarrow \nu_e$ probability is determined by the squared product of $U_{e4}$ and  $U_{\mu4}$~\cite{giunti_review} from Equation \ref{eq:PMNSMatrix}.  
This means that appearance results from LSND and MiniBooNE should be accompanied by $\nu_{\mu}$ and \nuebar disappearance.  
In fact, null $\nu_{\mu}$ disappearance results from IceCube and MINOS+ and null \nuebar disappearance results from reactor experiments rule out the phase space suggested by LSND and MiniBooNE~\cite{bib:kopp,bib:prl_joint2020}.  
Non-minimal 3+1 models can relax that incompatibility in a variety of ways.  
For example, an unstable sterile fourth mass state may decay into lighter $\nu_e$ or $\nu_{\mu}$, which can enhance $\nu_e$ appearance signatures or diminish $\nu_{\mu}$ disappearance signatures, respectively, while leaving other channels largely untouched~\cite{decay_conrad,decay_deg}.  
Some non-minimal sterile models can also relieve tension between interpretations of cosmology-oriented astrophysics datasets~\cite{cmb_s4} and terrestrial experiments~\cite{kopp_secret1, kopp_secret2,Song:2018zyl,nsi_denton}.  

While many non-minimal scenarios will produce signatures in PROSPECT-II consistent with a minimal 3+1 description, some will not.  
Considering the example above, the existence of a strongly-mixing (large $U_{e4}$) eV-scale sterile neutrino state that decays on longer timescales primarily to $\nu_{e}$ or $\nu_{\mu}$ would produce PROSPECT signatures equivalent to that of a minimal 3+1 model.  
In contrast, a weakly-mixing (small $U_{e4}$) sterile state with a fast decay to $\nu_{e}$ could result in a null \nuebar disappearance result in PROSPECT but visible $\nu_e$-like signatures in short-baseline accelerator experiments.  
In an alternate example of multiple eV-scale sterile neutrinos, PROSPECT's data can on its own elucidate the nature of the non-mimimal scenario~\cite{vsbl_mult}.  
These examples illustrate how PROSPECT's \nuebar disappearance datasets will make distinctive contributions to the global program to explore BSM possibilities involving sterile neutrinos.  

\begin{figure}[hptb!]
\centering
\includegraphics[trim = 0.0cm 12.0cm 15cm 0.0cm, clip=true, width=\textwidth]{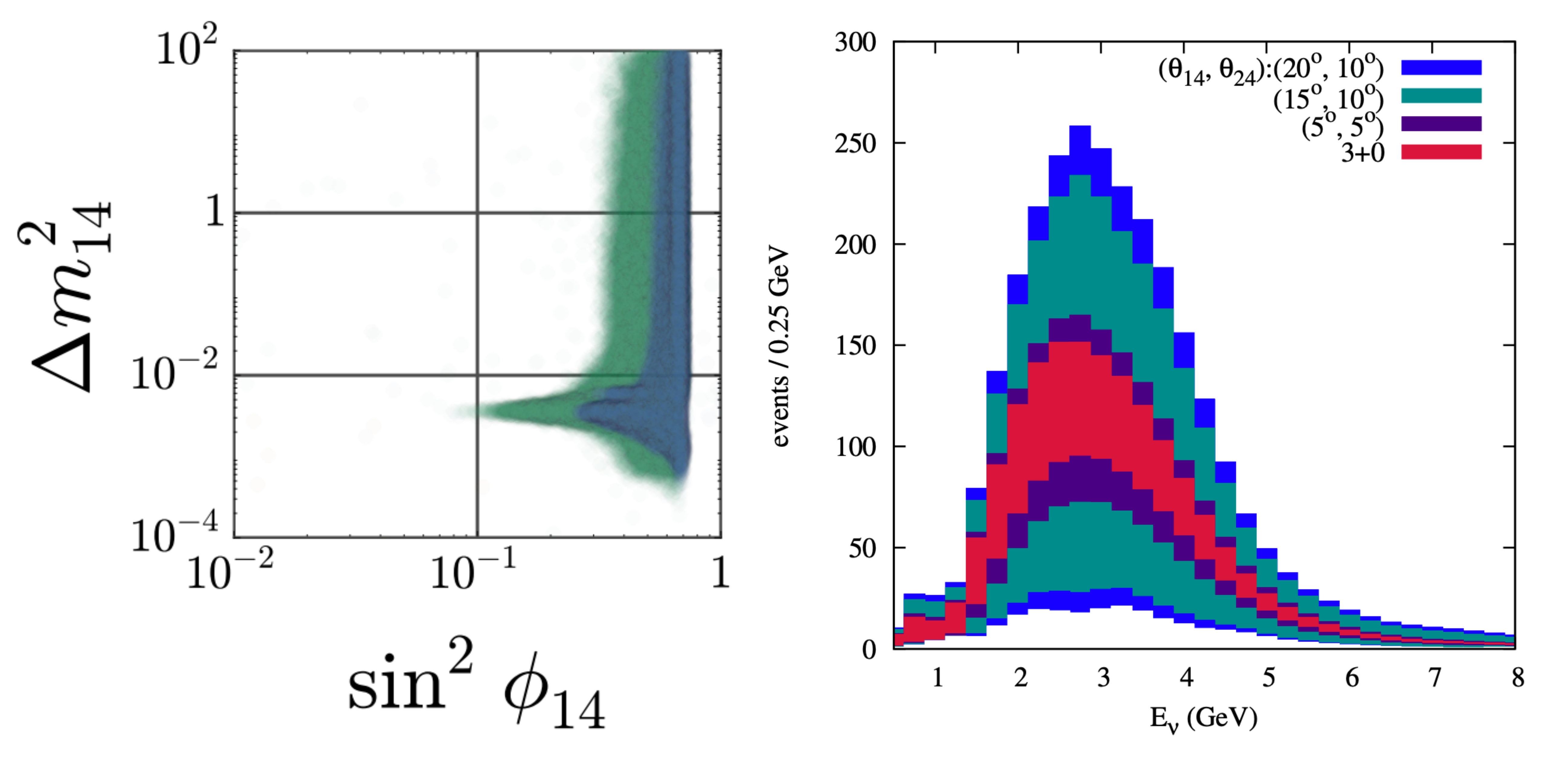}
\caption{
Left (from Ref.~\cite{cp_kelly}): values in $\theta_{14}$--$\Delta$m$^2_{41}$ space capable of generating false signals of CP invariance in long-baseline measurements despite a maximally CP violating value of $\delta_{13}$.  Note that $\phi_{14}=\theta_{14}$.  Right: (from Ref.~\cite{Kayser}): the range of observable DUNE $\nu_e$ appearance signals visible in the absence of sterile neutrinos (red) and in the presence of sterile states (others); the purple band indicates that the impact of sterile sector CP violation effects can be mitigated in DUNE if active-sterile mixing angles $\theta_{14}$ and $\theta_{24}$ are constrained to the level of 5$^{\circ}$ or less.  
}
\label{fig:cpv}
\end{figure}

Putting aside the goal of testing BSM physics models, one can also consider future short-baseline experiments in terms of the impact they have on measurement of Standard Model neutrino properties.
A number of studies have pointed to specific regions of 3+1 phase space, with no special relation to existing short-baseline anomalies, that could complicate the interpretation of future long-baseline neutrino measurements~\cite{Klop:2014ima,deGouvea:2014aoa}.  
A sterile neutrino with specific combinations of non-zero active and sterile CP violating phases $\delta_{13}$ and $\delta_{14}$ could mimic CP-conserved ($\delta_{13}$ = 0) signatures in DUNE \cite{Kayser}.  
Conversely, large-amplitude sterile mixing scenarios could yield false indications of CP violation in long-baseline experiments \cite{cp_kelly}.  
Illustrations of these parameter degeneracies are given in Figure~\ref{fig:cpv}.  

\begin{figure}[hptb!]
\centering
\includegraphics[trim = 0.0cm 0.0cm 0cm 0.0cm, clip=true, width=0.7\textwidth]{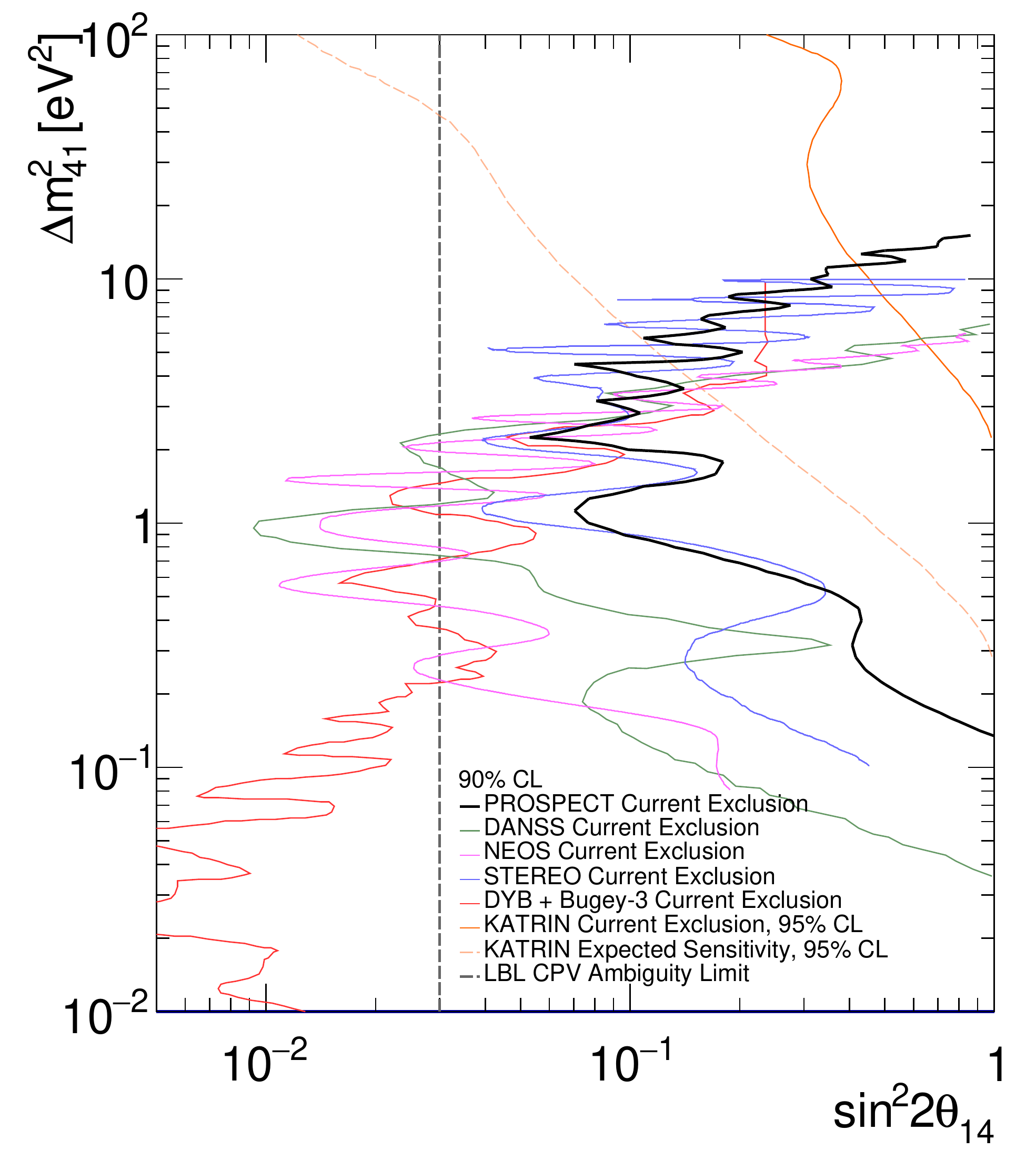}
\caption{
Sterile neutrino parameter space exclusion regions obtained by recent short-baseline reactor neutrino~\cite{bib:prl_joint2020,danss_osc,bib:neos,stereo_2019,prospect_prd} and tritium beta decay~\cite{katrin_sterile} experiments.  Cleaner interpretation of long-baseline oscillation results is possible if the active-sterile mixing amplitude in this parameter space is constrained to less than roughly 5$^{\circ}$, or sin$^2$2$\theta$ = 0.03 (black dashed line)~\cite{KayserVal}.
}
\label{fig:intro_cpvd}
\end{figure}

These parameter degeneracies can be avoided for future long-baseline experiments like DUNE if separate measurements can constrain the level of flavor mixing between the active and sterile sectors.  
In particular, it has been shown that these scenarios are avoided if limits on $\theta_{14}$ and $\theta_{24}$ can be improved to approximately the $5^{\circ}$ (sin$^2$2$\theta$ = 0.03) level~\cite{KayserVal}.  
While that limit has been achieved for $\theta_{24}$ and $\theta_{14}$ across many decades in mass splitting from $\nu_{\mu}$ and \nuebar disappearance experiments, respectively, it has not yet been met for $\theta_{14}$ over a mass splitting range from roughly 1 to more than 100 eV$^2$, as shown in Figure~\ref{fig:intro_cpvd}.  
Thus, future $\theta_{14}$-sensitive measurements, including PROSPECT-II, accelerator-based $\nu_e$ disappearance measurements~\cite{Abe:2014nuo,Abi:2020kei}, and tritium beta endpoint measurements~\cite{katrin_sterile}, will play a complementary role in interpretation of future long-baseline datasets.  

\subsection{Understanding the reactor antineutrino spectrum}
\label{subsec:intro_spec}

Alongside sterile neutrino searches, a motivation for PROSPECT has been to make a precise measurement of the neutrino spectrum from a highly enriched uranium reactor.  
Although it superficially resembles a falling exponential, this spectrum is the sum of thousands of unique, mostly short-lived beta decays and therefore depends on a very large set of nuclear data (cumulative fission yields for each daughter isotope, feeding strengths, etc.), forbiddenness for each beta transition, and a variety of higher-order correction factors~\cite{sonzogni_insights, vogel_review}.
For this reason, the reactor neutrino spectrum is difficult to precisely model.  
Spectrum models have advanced in quality since the 1980s, but both nuclear-data-based \emph{ab initio} and beta-spectrum conversion models continue to have difficulty replicating the observed spectrum \cite{vogel_review}.  

\begin{figure}[hptb!]
\centering
\includegraphics[trim = 0.0cm 14.0cm 12.0cm 0.0cm, clip=true, width=\textwidth]{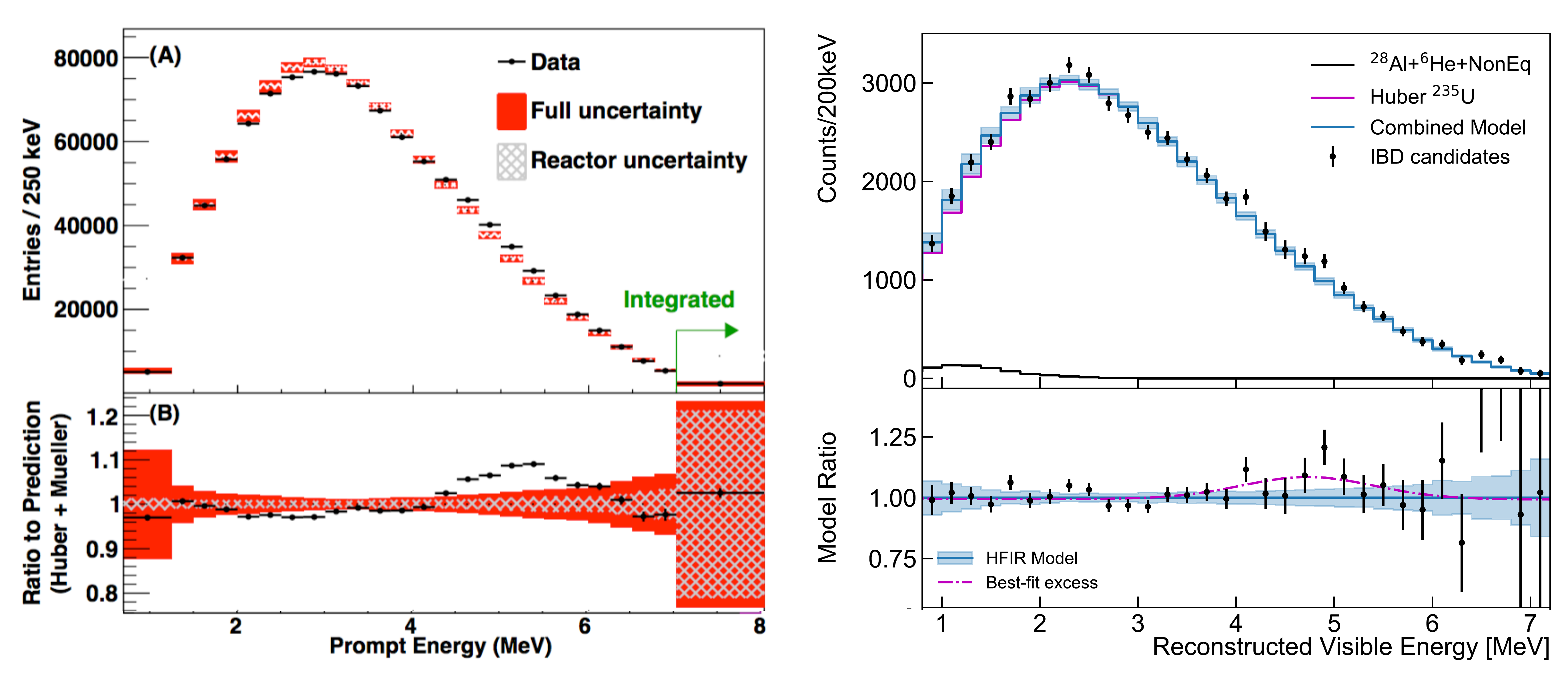}
\caption{Top panels: Observed and conversion-predicted IBD prompt spectra from the Daya Bay~\cite{bib:cpc_reactor} (left) and PROSPECT~\cite{prospect_prd} (right) experiments.  Bottom panels: ratio between predicted and measured spectra for both experiments.
}
\label{fig:intro_spec}
\end{figure}

An example of data-model discrepancy is an excess at high energies in data with respect to a similarly-normalized conversion-predicted spectrum.  
This disagreement, greatest in the 5-7~MeV region of \nuebar energy space, is often referred to as the `bump.'  
This feature was first observed in $\theta_{13}$  experiments in the mid-2010s~\cite{bib:reno_shape,dc_bump,bib:prl_reactor}.  
It has persisted in the new set of short-baseline reactor experiments including PROSPECT~\cite{prospect_spec} and STEREO~\cite{stereo_shape}.  
Data-model discrepancies observed in the Daya Bay and PROSPECT experiments are shown in Figure~\ref{fig:intro_spec}; the `bump' appears in these plots from 4-6~MeV in prompt energy space.  

Explaining the `bump’ has become a topic of scientific debate in the past 5-10 years among neutrino physicists and the nuclear science community.  
While it was initially suggested that this discrepancy was specific to the conversion approach~\cite{bib:dwyer}, investigations of \emph{ab initio} predictions using varied or updated nuclear databases showed that both prediction types suffer from a similar problem~\cite{hayes_shoulder, sonzongi2}. 
Re-measurement of beta decay datasets likely impacted by the pandemonium effect~\cite{tas_lots,tas_few,tas_br,tas_nb,tas_rb,tas_rbbr,tas_rbi} have improved \emph{ab initio} predictions; although improvements have reduced the flux discrepancy between \emph{ab initio} models and data, the `bump' has remained~\cite{bib:fallot2}.  
The spectral accordance between conversion and \emph{ab initio} models has directed focus towards common elements between these predictions, specifically the theoretical treatment of forbidden beta transitions~\cite{hayes_first,hayes_shoulder}.  
In particular, it has been suggested that mis-modeling of forbidden beta decay shape factors for high-yield, high-Q isotopes may be responsible for the shape anomaly~\cite{hayes_shape, hayen_initio}.  
For this reason, new nuclear physics measurements of spectrum shapes for individual isotopes have been highlighted for further study~\cite{bib:IAEA,osti_1649010}.  
It has also been suggested that the discrepancy may only be present in predictions for specific fission isotopes~\cite{hayes_shoulder,reno_evol}, despite the large overlap in contributing fission daughters between all isotopes~\cite{sonzogni_insights,Ma:2018fqf}.  

Neutrino experiments have also sought to identify the underlying issue through improved and diversified \nuebar measurements.  
The `bump' was first observed by experiments at LEU cores burning a combination of ~\uFive, \pNine, \uEight, and \pOne. As shown in Figure~\ref{fig:intro_spec}, PROSPECT, operating at the \uFive-burning HFIR core, showed a similarly-sized feature, indicating that \uFive~is at least partially responsible for the `bump' and that other isotopes likely contribute as well~\cite{prospect_prd}. A joint analysis from the PROSPECT and STEREO experiments has recently been released, strengthening this inference~\cite{bib:prosSTEREOjoint}. Daya Bay data collected during periods of different fuel content has yielded spectra of \uFive~and \pNine~\cite{bib:prl_235239}; these measurements similarly indicate that \uFive~is at least partially responsible for the `bump', while uncertainties in its \pNine~spectrum are still too large to provide a clear statement about other isotopes. A joint analysis of the PROSPECT and Daya Bay datasets has recently been released~\cite{bib:prosDBjoint}.  
A larger dataset from an HEU reactor will provide more insight into the the isotopic origin of the `bump,' with the potential to firmly establish \uFive~as a partial contributor, as Section \ref{subsec:phys_spec} describes.  
More generally, larger datasets enable precise inspection across the full  spectrum and possible identification of prediction issues in other energy ranges.  

Beyond its intrinsic scientific interest, a clearer picture of the reactor neutrino spectrum is now relevant to multiple projects in basic neutrino science and applied nuclear physics. On the science side, reactor neutrinos remain an important tool for precision measurements of Standard Model neutrino mixing parameters and the neutrino mass hierarchy~\cite{juno2}, as well as searches for BSM physics with reactor-based CEvNS measurements~\cite{Ricochet,Connie,conus}.  
For the latter case, IBD-based reference measurements will support precise, unbiased Standard Model predictions for future BSM CEvNS measurements.  
A resolution of data-model discrepancies in IBD prompt spectra will increase confidence in model predictions at very high and very low reactor neutrino energies, which are not accessible to IBD-based experiments.  

On the applications side, reactor neutrino spectra, as well as energy-integrated fluxes, have two points of relevance. 
First, the nuclear data community maintains detailed databases of the properties of individual nuclear fission products for use in a variety of nuclear energy and nuclear security programs. 
Precise measurements of the neutrino spectrum generated by these fission products represents a novel nuclear data tool to validate those databases.  
This concept of `neutrino spectroscopy'~\cite{sonzogni_fine} inverts the traditional frame of modeling reactor neutrino spectra from nuclear data, instead using neutrino data to validate nuclear measurements.  
This approach offers advantages with respect to tools currently in the nuclear data pipeline: current criticality validations~\cite{crit} primarily use datasets relating to the neutronics of nuclear interactions, such as fission neutron yields and neutron absorption properties, while neutrino-based validations check the fidelity of data related to the production and decay properties of fission products.  

Additionally, the possibility of using neutrino detectors themselves in nuclear security has gained some attention in recent years~\cite{Bernstein:2019hix}. Proposed concepts include using neutrino detectors as a safeguards tool for advanced reactor designs or as a verification element in future nuclear agreements. Meeting the exacting standards of these applications will require firm understanding of the reactor source term, which PROSPECT-II will advance through the flux and spectrum measurements described in Section \ref{subsec:phys_flux} and \ref{subsec:phys_spec}. PROSPECT-II will also demonstrate technology features relevant to cooperative verification and safeguards applications. As in the first phase of PROSPECT, the PROSPECT-II detector will be largely assembled offsite and installed in an aboveground location at HFIR (with overburden $<1$ m water equivalent). Once installed, the detector is expected to operate with minimal intervention for a two-year period. The design of the detector will support movement to another reactor site for further measurements.

%% file: PII_PhysicsGoals.tex
The PROSPECT collaboration is now preparing an evolutionary upgrade of its detector, called PROSPECT-II, that leverages existing investments and expertise, resolves technical issues that abbreviated the first run, introduces design features that improve robustness and time-stability, and extends both the depth and the scope of the experiment’s physics reach.  
The upgraded PROSPECT-II detector will provide world-leading oscillation and \uFive~\nuebar spectrum sensitivity, produce a new measurement of the total \uFive~\nuebar yield, and enable a future expanded physics program via relocation to a commercial LEU reactor.  
The primary goals of PROSPECT-II are:

\begin{itemize}
    \item Searching for mixing between active and sterile neutrino sectors in the mass-splitting range of $\sim$1-20~eV$^2$, covering a region beyond the reach of other reactor experiments;  
    \item Extending sensitivity to the sterile mixing angle $\theta_{14}$ below $\sim$5$^{\circ}$ to inform the interpretation of long-baseline CP violation experiments;
    \item Reducing $^{235}$U spectrum uncertainties below 5\,\%, uniquely constraining reactor predictions;
     \item Performing an absolute measurement of the $^{235}$U \nuebar{} yield and improving the robustness of the global yield picture for the three dominant fission isotopes \uFive, \pNine, and \uEight;
     \item Enabling a future measurement program with highly correlated detector systematics at a commercial reactor to expand the covered $\Delta m_{41}^2$ range and strengthen spectrum and flux constraints.  
\end{itemize}

The following sections summarize the design and achievements of PROSPECT's first run, overview the fundamental design features of PROSPECT-II, and present the \nuebar oscillation, spectrum, and flux physics goals of PROSPECT-II.  

\subsection{Review of the first phase of PROSPECT}
\label{subsec:p1}

PROSPECT began its first run in 2018 at the High Flux Isotope Reactor (HFIR) at Oak Ridge National Laboratory. The active \nuebar detector, PROSPECT-I, was a segmented 4-ton $^6$Li-doped, PSD-capable liquid scintillator volume optimized for inverse beta decay detection with minimal cosmic-ray shielding. The PROSPECT-I detector is described in greater detail in Section \ref{sec:PI_Detector} and Refs.~\cite{prospect_nim,prospect_prd},  

Data from the first run of PROSPECT, comprising 183 calendar days, set new limits on the existence of eV-scale sterile neutrinos in the region preferred by global flux data \cite{prospect_prd}, as illustrated in Figure~\ref{fig:current_osc}. 
The first PROSPECT run also provided the world’s most precise \nuebar energy spectrum produced by a highly-enriched $^{235}$U reactor~\cite{prospect_prd}, as illustrated in Figure~\ref{fig:intro_spec}.  
The broader impact of these measurements was discussed in detail in the previous section.  
Beyond these flagship measurements, PROSPECT also provided new estimates of non-fuel \nuebar production by research reactors~\cite{nonfuel} and set new limits on low-mass WIMP dark matter~\cite{prospect_dm}.   

In addition to these scientific achievements, PROSPECT made notable technical progress by successfully operating a neutrino detector at the earth's surface capable of achieving a signal-to-cosmic-background ratio of better than 1:1.  
PROSPECT is the first experiment to have demonstrated this capability, putting it at the forefront of global scientific efforts to develop \nuebar detectors for reactor monitoring applications.  
The PROSPECT-I detector also achieved a ratio of IBD signal to accidental backgrounds of better than 1:1 in a challenging reactor-adjacent  environment.  
These background reduction achievements have validated the suitability of PROSPECT's HFIR deployment location for \nuebar science, the efficacy of the existing PROSPECT passive shielding package, and the rejection power delivered by the combination of segmentation, $^6$Li-doping, and IBD prompt and delayed signal PSD inherent in the PROSPECT-I active detector design.

PROSPECT achieved these benchmarks despite curtailed operation due to a combination of detector technical issues and unexpected HFIR downtime that limited data-taking to a one-year timeframe. The gradual loss of function of some detector segments, as described in Section \ref{sec:PI_Detector}, has been addressed through analysis adaptations. The oscillation search and spectrum measurement results are, at present, limited by statistical uncertainty on the relatively small dataset collected in the first PROSPECT run.

\subsection{PROSPECT-II design and experimental parameter overview}
\label{subsec:phys_params}

The original PROSPECT experimental siting at HFIR and detector design initially met all installation and performance requirements.  
Thus, PROSPECT-II builds upon the original PROSPECT design by making evolutionary changes to the active detector to realize a more robust system capable of multi-year operation at one or more sites.  
While unchanged and upgraded design aspects will be discussed in detail in Section~\ref{sec:PII_Detector}, this section highlights a few improved features of PROSPECT-II particularly relevant to its expected physics capabilities:

\begin{itemize} 
 \item \emph{Movement of PMTs outside the scintillator volume}: ensures long-term functionality of all detector segments, increasing active detector volume and reducing backgrounds.  
 \item \emph{Extended data collection periods, 25\% longer segments,  and 20\% increased LS $^6$Li doping fraction:} increases the total number of detectable IBD interactions in PROSPECT-II.  
\item \emph{Reduction in number of different materials in contact with scintillator and new in-situ scintillator replacement capability:} ensures PROSPECT-II's long-term stable operation and enables movement to other reactor sites.  
\item \emph{Adjusted calibration design elements and assembly/filling QA procedures}: enables improved determination of absolute detection efficiency, fiducial target mass, and absolute energy scale.  
\end{itemize}

Table~\ref{tab:exp_param} summarizes nominal experimental parameters for a PROSPECT-II deployment at HFIR and at a commercial LEU reactor, alongside realized parameters for the first PROSPECT run~\cite{prospect_prd}.  
The detector and exposure parameters in Table~\ref{tab:exp_param} convey the level of similarity in nominal parameters between PROSPECT-II and the first PROSPECT run.  
As described in the list above, primary differences between the two include total deployment duration and IBD statistics at HFIR, segment length, and expected background contributions.  
Through the remainder of this section, these parameter sets are used as inputs for discussion of the oscillation, spectrum, and flux physics capabilities of the PROSPECT-II detector.  

\begin{table*}[thbp!]
\centering
\begin{tabular}{c|l|c|c|c}
\hline
\multicolumn{2}{c|}{Parameter} & P1 & P2 at HFIR & P2 at LEU  \\ \hline 
\multirow{4}{*}{Reactor} & Power (MW$_\textrm{th}$) & \multicolumn{2}{|c|}{85} & 3000 \\ 
& Cylinder Size ($d\times$h, m$^2$) & \multicolumn{2}{|c|}{$0.4 \times 0.5$} & $3 \times 3$ \\ 
& Fuel & \multicolumn{2}{|c|}{HEU} & LEU \\ 
& Cycle Length & \multicolumn{2}{|c|}{24~d} & 1.5~y\\ \hline \hline
\multirow{7}{*}{Detector} & Segmentation & 11$\times$14 & \multicolumn{2}{|c}{11$\times$14} \\ 
& Segment Area (cm$^2$) &$14.5\times14.5$ & \multicolumn{2}{|c}{$14.5\times14.5$} \\ 
& Segment Length (m) & 1.17 & \multicolumn{2}{|c}{1.45} \\ 
& Target Mass (ton) & $\sim$4.0 & \multicolumn{2}{|c}{4.8} \\ 
& Light collection (PE/MeV) & $\sim$380 & \multicolumn{2}{|c}{500} \\ 
& Detection Efficiency & $\sim$40\% & \multicolumn{2}{|c}{40\%} \\ \hline \hline
\multirow{6}{*}{Exposure} & Average Baseline (m) & 7.9 & 7.9 & 25 \\ 
& Reactor-On Days (d) & 105 & 336 & 548 \\ 
& Reactor-Off Days (d) & 78 & 360 & 61 \\ 
& Signal:Background & 1.4 & 4.3 & 19.3 \\ 
& IBD Statistics ($N_{IBD}$) & 50560 & $3.74 \times 10^5$ & $2.72 \times 10^6$ \\
& Effective Statistics ($N_{\textit{eff}}$) & 15195 & $2.08 \times 10^5$ & $1.79 \times 10^6$ \\
\hline
\end{tabular}
\caption{Experimental parameters for the first run of PROSPECT (`P1'), as well as nominal parameters for PROSPECT-II (`P2') deployments at HFIR and at a commercial LEU reactor.  Realized parameters for PROSPECT's first run are taken from Ref.~\cite{prospect_prd}.  
}
\label{tab:exp_param}
\end{table*}

Deployment of an upgraded PROSPECT-II detector for a multi-year data-taking run at HFIR will enable a significant improvement in IBD signal statistics, which are currently the principal limitation of PROSPECT's centerpiece oscillation and spectrum measurements.  
In PROSPECT, the total number of background-subtracted IBD signals (`IBD Statistics' in Table~\ref{tab:exp_param}) is defined by the detected IBD-like candidate datasets as:
\begin{equation}
\label{eq:ibd_stats}
N_{IBD} = N_{\textrm{On}} - N_{\textrm{On},a} - R(N_{\textrm{Off}} - N_{\textrm{Off},a}),
\end{equation}
where $N_{\textrm{On}}$ and $N_{\textrm{On,a}}$ are the detected IBD-like candidates and estimated accidental backgrounds during reactor-on periods, $N_{\textrm{Off}}$ and $N_{\textrm{Off},a}$ are the same but for reactor-off periods, and $R$ is the relative on-to-off live-time ratio.
The PROSPECT-II detector at HFIR will be capable of recording roughly 1100 IBD per day, amounting to more than 350,000 over 14 reactor cycles (2 calendar years).

More relevant to the primary physics analyses is the combined statistical uncertainty of both signal and background, which can be expressed in terms of the size of a background-free IBD dataset generating a similar level of relative statistical uncertainty, denoted `Effective Statistics' in Table~\ref{tab:exp_param}: 
\begin{equation}
\label{eq:eff_stats}
N_{\textit{eff}} = \frac{((N_{\textrm{On}}-N_{\textrm{On},a})-R(N_{\textrm{Off}}-N_{\textrm{Off},a}))^{2}}{(N_{\textrm{On}}+N_{\textrm{On},a})+R^{2}(N_{\textrm{Off}}+N_{\textrm{Off},a})}
\end{equation}
With an expected signal-to-background ratio of 4.3 in the PROSPECT-II detector, a 2-year PROSPECT-II run at HFIR will provide $>$200,000 effective IBD counts; this represents more than an order of magnitude improvement beyond the roughly 15,000 effective IBD counts available for PROSPECT's most recent results~\cite{prospect_prd}.  
As shown in Table~\ref{tab:exp_param}, effective statistics for an LEU-based deployment of PROSPECT-II could be expected to exceed those of the improved HEU-based deployment by roughly an additional order of magnitude.  

The remainder of this section describes how the PROSPECT-II dataset can be used to extend PROSPECT's sterile oscillation and \uFive~neutrino energy spectrum physics impact.  
This section also describes how improvements in the PROSPECT-II detector's systematic uncertainties and robustness will enable a modern~\uFive~neutrino flux measurement and situate the experiment for an expanded long-term oscillation and spectrum physics program.  

\subsection{PROSPECT-II oscillation physics goals}
\label{subsec:phys_osc}

The main particle physics goal of PROSPECT-II is to pursue the physics associated with neutrino mass by searching for coupling between active and sterile neutrino sectors~\cite{prospect}.  
A detailed discussion of the motivations for this measurement can be found in Section~\ref{sec:Introduction}.
The compact HFIR core and very short reactor-to-detector baseline will allow PROSPECT-II to continue pushing sensitivity in the mass splitting range above a few eV$^2$, where the first run of PROSPECT was most sensitive.  
The following sub-section focuses on oscillation sensitivity gains from the PROSPECT-II deployment at HFIR; gains from future deployments beyond HFIR are described in Section~\ref{subsec:phys_mult}

To test for the presence of neutrino oscillations in PROSPECT, the collaboration performs relative comparisons of observed IBD prompt energy spectra between different baseline ranges in the detector.  
Each prompt spectrum consists of IBD candidates with interaction locations reconstructed within a subset of detector segments of common average reactor-detector baseline.
Spectrum ratios between baselines can then be compared to predicted ratios in the presence of varying oscillation signatures dictated by the parameters ($\Delta$m$^2_{41}$, sin$^22\theta_{14}$) using a $\chi^2$ test statistic.  
Predicted ratios include the impact of modest expected detector response differences between baselines.  
The impact of both correlated and uncorrelated systematic and statistical uncertainties are included in the $\chi^2$ test statistic using a covariance matrix approach.  
This sterile neutrino fitting procedure, as well as the input collected and simulated datasets used for this analysis in PROSPECT's first run, are described in detail in Ref.~\cite{prospect_prd}.  

Asimov datasets are generated with the same tools used to generate oscillated spectrum predictions, with reactor fluxes modeled using Ref.~\cite{bib:huber} and the IBD interaction cross-section modeled using Ref.~\cite{Vogel:1999zy}.
Asimov datasets are generated with the same tools used to generate oscillated spectrum predictions.  
For PROSPECT-II estimates, detector response effects are re-simulated and all systematic uncertainty covariance matrices are re-calculated to reflect the expected functionality of all detector segments.  
Other systematic uncertainties established using PROSPECT-I data, such as those related to linear and non-linear energy scales, baselines, relative efficiency differences, and background normalization, are used for PROSPECT-II.  
Sensitivity estimates are made using ten baseline bins; a more finely binned analysis may produce modestly improved coverage~\cite{VSBL, pros_stats}.  

\begin{figure}[hptb!]
\centering
\includegraphics[trim = 0.0cm 0.0cm 0.5cm 0.0cm, clip=true, width=0.475\textwidth]{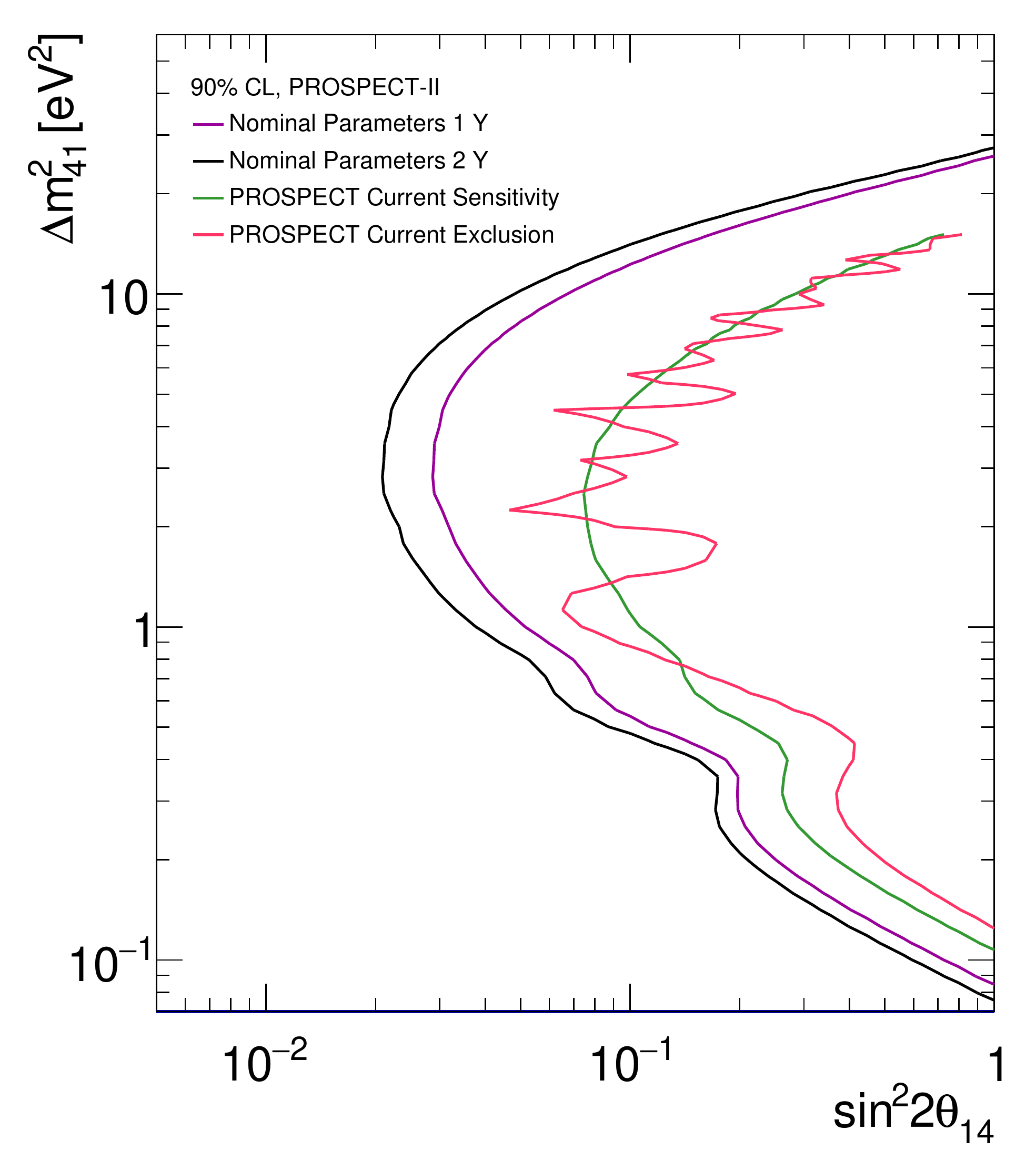} \\
\includegraphics[trim = 0cm 0.0cm 0.5cm 0.0cm, clip=true, width=0.475\textwidth]{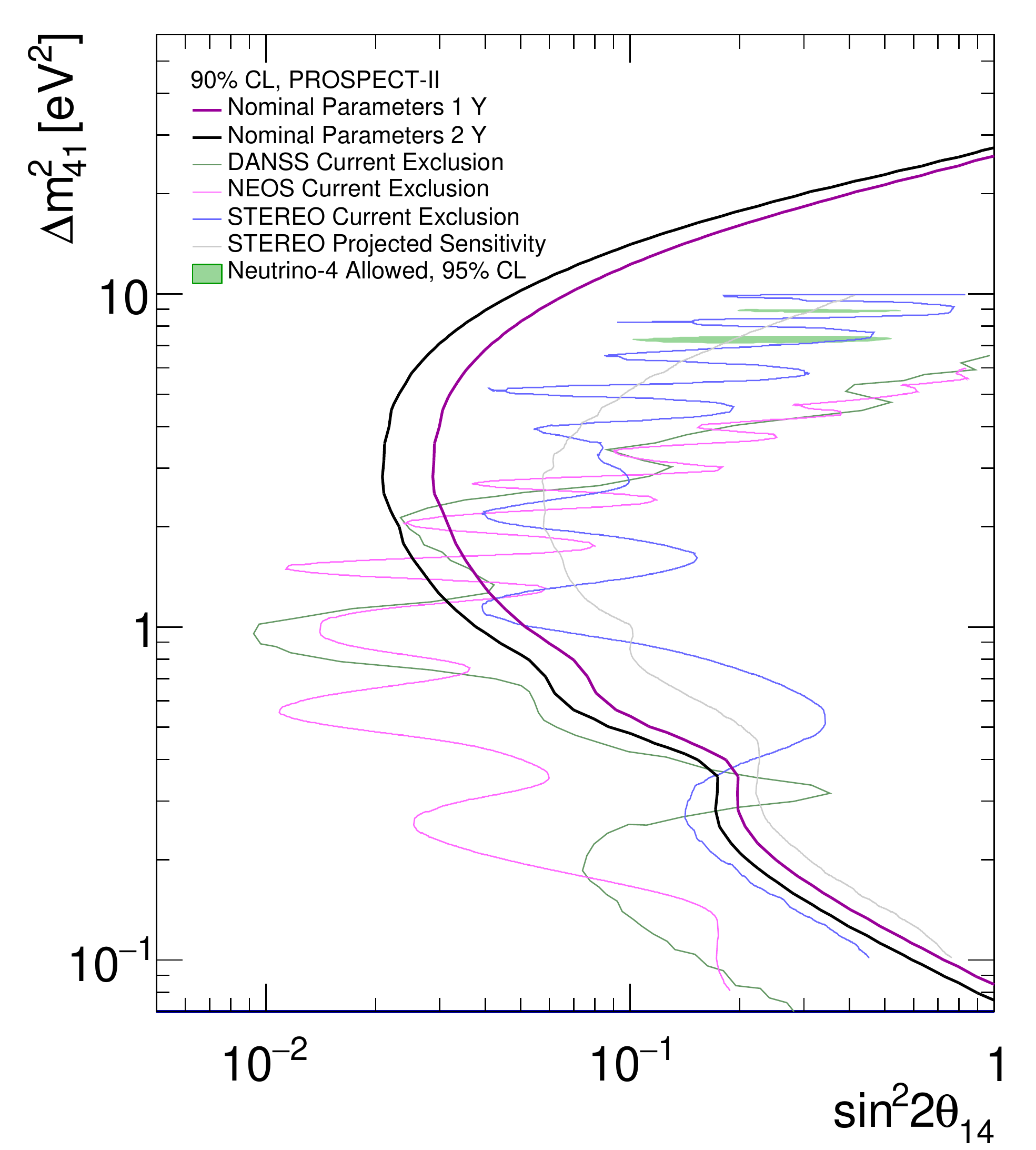}
\includegraphics[trim = 3.0cm 0.0cm 0.5cm 0.0cm, clip=true, width=0.404\textwidth]{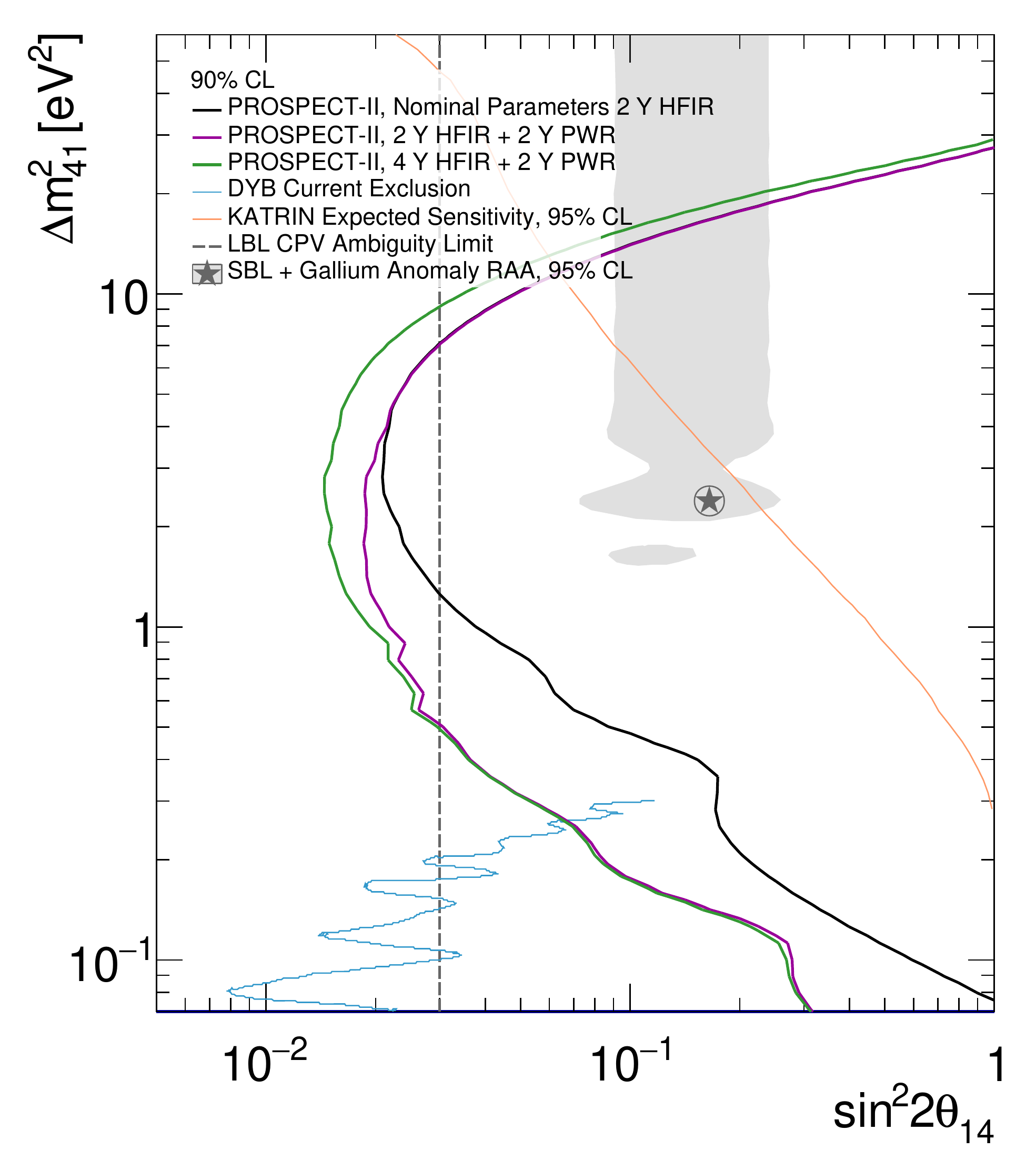}
\caption{Top: Comparison of sterile oscillation sensitivities for different current and projected PROSPECT~\cite{prospect_prd} and PROSPECT-II datasets.  Bottom left and right: Overlap of sensitivity contours from one year to two years of PROSPECT-II data-taking with various interesting regions of oscillation parameter space, as described in the text. The scenarios in the right plot including LEU data are described in Section \ref{subsec:phys_mult}. Non-PROSPECT contours from Refs.~\cite{danss_osc,bib:neos,stereo_2019,bib:neutrino4_osc,bib:prl_joint2020,katrin_sterile,giunti_diagnose}. Expected sensitivities are shown for the two experiments that have published these projections, STEREO and KATRIN (shown in bottom left and right plots, respectively).}
\label{fig:osc}
\end{figure}

Projected sensitivities for PROSPECT-II using the nominal parameters from Table~\ref{tab:exp_param} are shown in Figure~\ref{fig:osc}.  
Current PROSPECT oscillation results, also shown in Figure~\ref{fig:osc}, are primarily statistics-limited.  
After two calendar years of data-taking at HFIR, PROSPECT-II will improve upon the sensitivity of the most recent PROSPECT measurement above 0.5~eV$^2$ by a factor of anywhere from 
two to six, depending on the mass splitting.  

The middle panel of Figure~\ref{fig:osc} provides a view of PROSPECT-II's projected sensitivity in context with current and future results from other reactor short-baseline experiments.  
After a two-year run at HFIR, PROSPECT-II will provide the best coverage of any short-baseline reactor experiment at all mass splitting values above $\sim$1.5~eV$^2$.  
The compact-core-based STEREO experiment, while providing good sensitivity in this high mass  region~\cite{stereo_2019}, has limited statistical precision and has no plans for future upgrades or expanded data-taking. PROSPECT-II's sensitivity will exceed that of the final expected STEREO result by more than a factor of three.  
The only other current or planned experiment with substantial sensitivity in this high mass range, Neutrino-4, has, as noted in Section \ref{sec:Introduction}, controversially claimed an observation of oscillations near 7~eV$^2$~\cite{bib:neutrino4_osc} in spite of unaddressed analysis and statistical issues that have been repeatedly raised by the community~\cite{n4_comment}.  
Nonetheless, broader reception of the result has been mixed, leaving Neutrino-4's claim as a remaining open question.  
As Figure~\ref{fig:osc} shows, in  a single year, PROSPECT-II can conclusively address the Neutrino-4 claim.  

The right panel of Figure~\ref{fig:osc} places PROSPECT-II sensitivities in the context of other relevant excluded, suggested, or interesting regions of parameter space.  
Within two years of data-taking at HFIR, PROSPECT-II will entirely address the remaining parameter space highlighted by the oscillation-based explanation for the Reactor Antineutrino Anomaly up to 15~eV$^2$.  
The exclusion of phase space above PROSPECT's measurement limits will be provided by KATRIN's future tritium beta endpoint measurements~\cite{katrin_sterile}.  
Thus, the addition of PROSPECT-II to the full set of complementary global constraints can enable resolution of this long-standing source of interest in the neutrino physics community.  

In addition to being compelling in its own right, PROSPECT-II's sensitivity to active-sterile mixing will support the broader global effort to understand the physics associated with neutrino mass.  
As mentioned in Section~\ref{subsec:sterile_broad}, sizable couplings between sterile and active neutrino sectors in the mass splitting regime above $\Delta$m$^2_{32}$ can obscure interpretation of long-baseline CP-violation experimental results.  
As illustrated in Figure~\ref{fig:osc}, PROSPECT-II's coverage will address an important gap in sensitivity that would otherwise exist between highly-sensitive Daya Bay disappearance results at low $\Delta$m$^2$ ($<$0.2~eV$^2$)~\cite{bib:prl_sterile} and future KATRIN beta-endpoint measurement results at high $\Delta$m$^2$ ($>$10 eV$^2$).  
With two years of data-taking at HFIR, PROSPECT can achieve 95\% confidence level active-sterile mixing precision exceeding the sin$^22\theta_{14}=0.03$ guideline advised for clarifying long-baseline interpretations~\cite{KayserVal} over much of the parameter space below 10~eV$^2$.  
Due to the very large size of commercial LEU reactor cores, the DANSS~\cite{danss_osc}, NEOS~\cite{bib:neos}, and JUNO-TAO~\cite{Berryman:2021xsi} short-baseline reactor experiments, even with projected future data-taking, will not be able to sensitively probe oscillations above a few~eV.  

\subsection{PROSPECT-II reactor \nuebar spectrum physics goals}
\label{subsec:phys_spec}

Deployment of the PROSPECT-II detector at HFIR will also improve the precision of PROSPECT's world-leading measurement of the $^{235}$U $\overline{\nu}_e$ energy spectrum.   
Currently, comparisons of reported \nuebar energy spectra between theoretical predictions and experimental results obtained at LEU reactors show sizable disagreements, most obviously in the higher-energy 4-6~MeV reconstructed energy range~\cite{bib:prl_reactor,bib:reno_shape,dc_bump}.  
Possible explanations for the source of this discrepancy are discussed in detail in Sec~\ref{subsec:intro_spec}, along with the general value of precise understanding of all aspects of fission-produced \nuebar spectra.  
The PROSPECT measurement of \nuebar energies from~\uFive~contributes to this understanding by helping to determine whether or not current experiment-theory disagreements are common across fission isotopes.  

\begin{figure}[hptb!]
\centering
\includegraphics[trim = 0.5cm 0.5cm 0.5cm 0.5cm, clip=true, width=0.49\textwidth]{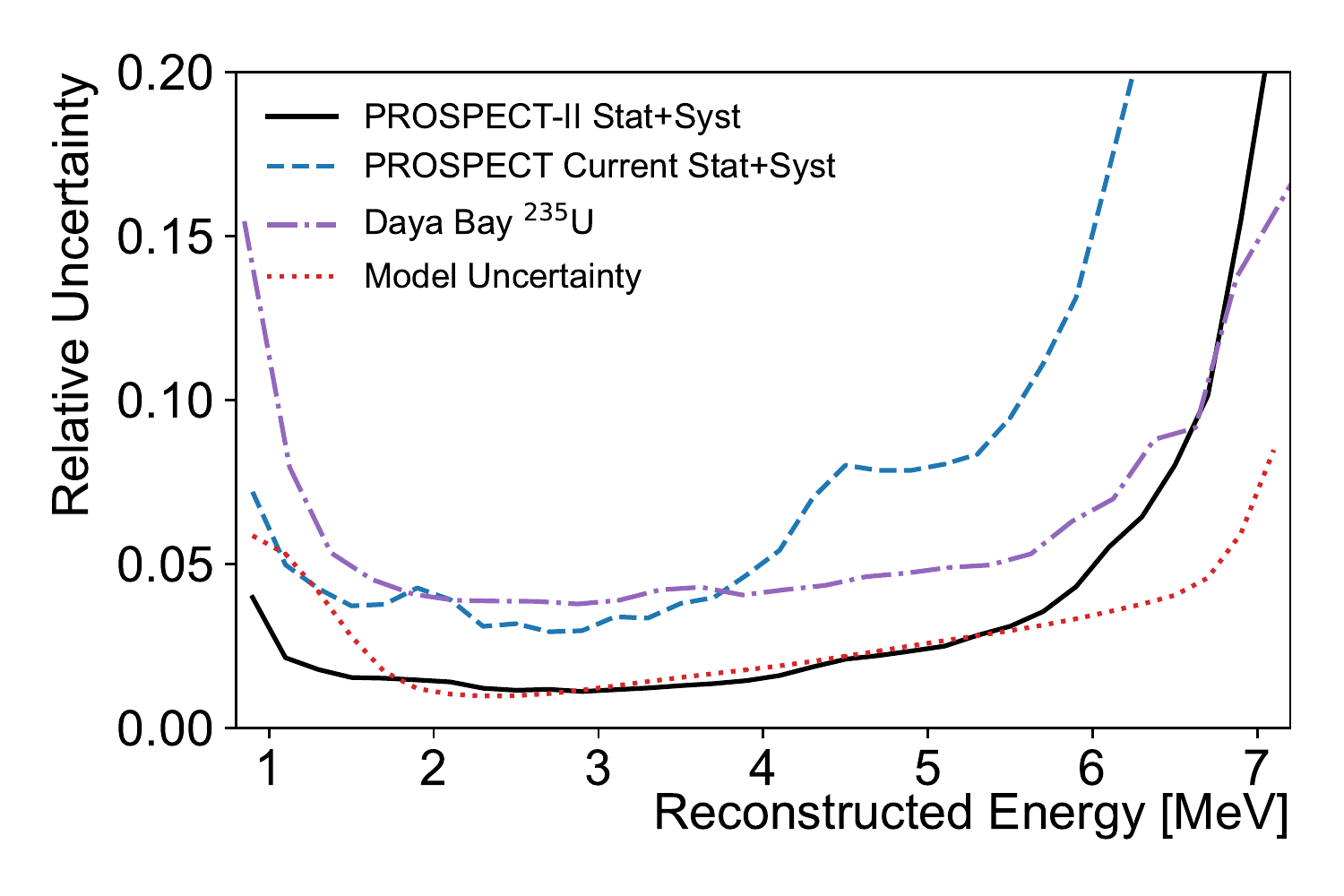}~
\includegraphics[trim = 0.5cm 0.5cm 8.3cm 0.5cm, clip=true, width=0.49\textwidth]{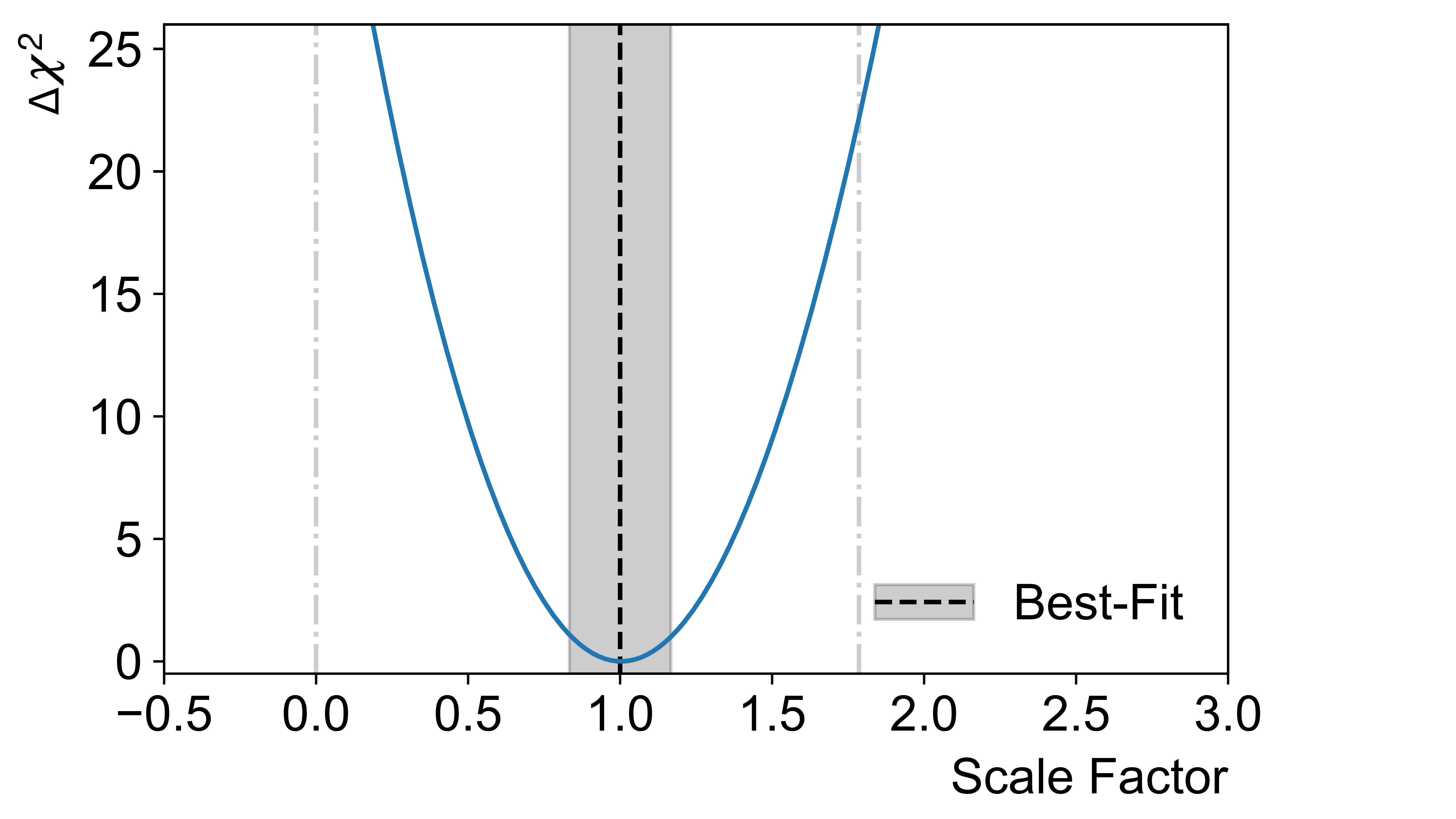}
\caption{Left: Improvement to PROSPECT $^{235}$U spectrum measurement uncertainties after two years of data-taking with the PROSPECT-II detector.  The depicted model uncertainty, from Ref.~\cite{prospect_prd}, is based on the conversion prediction of Ref.~\cite{bib:huber}. Also shown is the \uFive{} uncertainty from deconvolution of LEU measurements at Daya Bay from~\cite{Adey:2021rty}.  
Right: PROSPECT-II  measurement precision for the the scale factor (n) of a bump-like feature in the 4-6~MeV prompt energy region observed by Daya Bay.  PROSPECT will address prominent hypotheses for this feature (n=0: no contribution from \uFive; n=1.78: entirely from \uFive) at high confidence level.}
\label{fig:spectrum}
\end{figure}

The impact of a two-year PROSPECT-II measurement at HFIR is illustrated in Figure~\ref{fig:spectrum}.  
Additional reactor-on running, plus a reduction in backgrounds with respect to PROSPECT's first run (reflected in Table~\ref{tab:exp_param}) due to the full functionality of all detector segments, will substantially reduce statistical uncertainties on the \uFive~\nuebar measurement.  
Statisical uncertainties per 200~keV bin will be reduced below the 3\% level in all regions below 6~MeV in prompt energy, roughly 7~MeV in \nuebar energy, and will be as low as $\sim$1\% at the spectrum peak.  
Precision in the \uFive~spectrum from PROSPECT-II's HEU-based measurement will exceed that achievable in LEU-based `evolution' measurements taking advantage of changes in fuel content, such as Daya Bay~\cite{Adey:2021rty}, which is also illustrated in~Figure~\ref{fig:spectrum}.  
It should be noted that future LEU-based short-baseline experiments, such as JUNO-TAO~\cite{juno_tao} and NEOS-II~\cite{neos2}, are unlikely to exceed the statistical precision of the current Daya Bay dataset, making it difficult to 
improve upon the LEU-derived \uFive~error bars reported in Figure~\ref{fig:spectrum}.  
Moreover, PROSPECT-II’s HEU-based measurement is unaffected by uncertainties in fuel content evolution and in the \nuebar spectra of sub-dominant isotopes present in LEU cores.  

\begin{figure}
    \centering
    \includegraphics[width=0.6\textwidth]{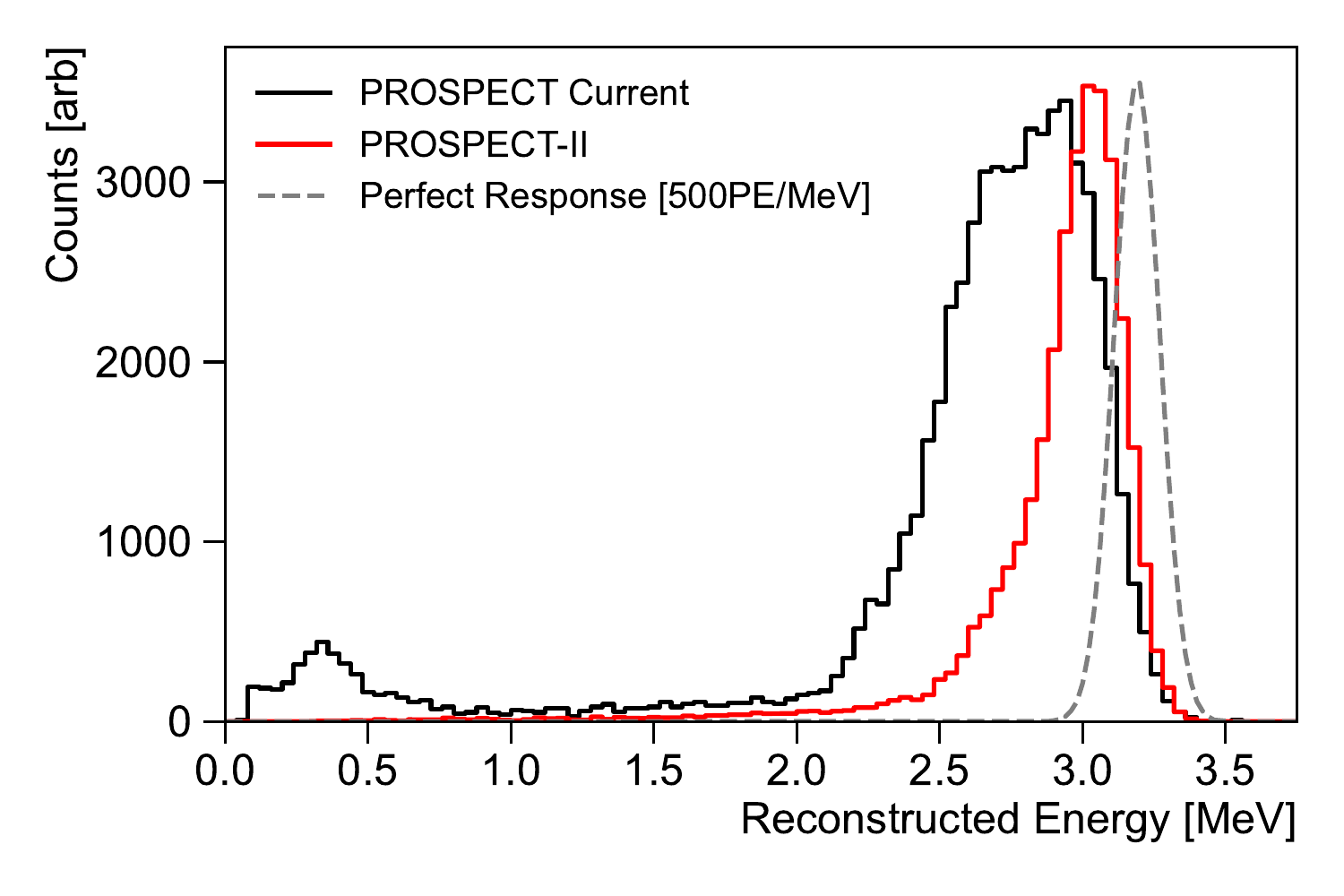}
    \caption{Comparison of simulated detector response to monoenergetic 4~MeV \nuebar for the PROSPECT-II detector (red) and the most recent PROSPECT publication~\cite{prospect_prd} (black).  For comparison, the gray dashed line shows a Gaussian representing the response of a perfect detector with no energy loss in non-scintillating materials and a width matching that expected from 500 PE/MeV scintillation light collection.}
    \label{fig:detector_response}
\end{figure}

Relative to PROSPECT-I, detector systematic uncertainties will be reduced and effective neutrino energy resolution improved through the elimination of lost IBD positron energy in non-functioning detector segments.  
Figure~\ref{fig:detector_response} illustrates these expected improvements in detector response by comparing simulated detector response to a monoenergetic source of 4~MeV \nuebar in the latest PROSPECT analysis~\cite{prospect_prd} to the projected PROSPECT-II response.  
The reconstructed energy peak in PROSPECT-II exhibits a full width at half maximum less than half of that realized in existing PROSPECT results. PROSPECT-II's reconstructed energy bias with respect to the depicted full-energy peak is also reduced by more than a factor of two.  

Incorporating the combined statistical and detector response improvements described above, PROSPECT-II will provide a \uFive~spectrum measurement significantly improved from recently-reported PROSPECT measurements, as shown in Figure~\ref{fig:spectrum}~\cite{prospect_prd}.  
It is expected that a two-year run using the existing PROSPECT event selection will result in a measurement that approaches or exceeds the precision of beta-conversion and summation predictions and that is unlikely to be matched by other HEU short-baseline experiments.  
When jointly analyzed with spectrum measurements from Daya Bay, PROSPECT-II's measurement can be used to produce purely data-driven reactor $\overline{\nu}_e$ spectrum models for future particle physics measurements and potential applications.

Beyond improving the overall precision on global knowledge of the $^{235}$U~\nuebar energy spectrum, PROSPECT-II can aid in better understanding the limitations and features of existing reactor \nuebar models.  
In particular, PROSPECT-II will provide 
new information about the nature of the 4-6~MeV spectrum `bump' anomaly, as shown in Figure~\ref{fig:spectrum}.  
The PROSPECT measurement of this feature's amplitude in $^{235}$U relative to Daya Bay's will improve in precision from the current 39\%~\cite{prospect_prd} to $\sim$15\%. 
In the event that PROSPECT-II observes a bump size identical to Daya Bay, PROSPECT would be able to exclude $^{235}$U as being a non-contributor (the sole contributor) to this feature with $>$5$\sigma$ ($>$4$\sigma$) confidence, definitively addressing a point of interest in the reactor neutrino and nuclear data communities.  
A high-statistics PROSPECT-II measurement will also be a means to investigate other features in \textit{ab initio} $^{235}$U spectrum predictions made from nuclear databases~\cite{bib:dwyer}, such as deviations from conversion predictions at very low and very high energies. 

\subsection{PROSPECT-II reactor \nuebar flux physics goals}
\label{subsec:phys_flux}

A data-taking campaign with the PROSPECT-II detector at HFIR will allow a precise measurement of the $\overline{\nu}_e$ flux produced in $^{235}$U fission.   
As described in Section \ref{sec:Introduction}, historical measurements of HEU and LEU \nuebar fluxes~\cite{bib:mention2011}, as well as modern LEU flux measurements~\cite{bib:prd_rate,reno_evol,dc_nature}, show a net flux deficit with respect to beta-conversion models~\cite{giunti_diagnose,huber_berryman}.
At LEU reactors, the emitted \nuebar flux evolves over time as the isotopic composition changes in the fuel cycle, a measured effect~\cite{bib:prl_evol,reno_evol} that is also improperly predicted by existing beta-conversion models.  
Further analysis of these results suggest many possible causes, which were discussed in detail in Section~\ref{subsec:raa_flux}.  
Currently, global fits that attempt to understand this effect are strongly influenced by HEU flux measurements that are more than 30 years old; many of these measurements have already received focused criticism in the literature~\cite{ILL_fix, bib:chao}.  

By performing a modern \uFive~\nuebar flux measurement, PROSPECT-II can increase the reliability of the global flux picture,
benefiting both the particle physics and nuclear science communities in ways similar to those outlined in Section~\ref{subsec:intro_spec}.  
A PROSPECT-II flux measurement can also provide an independent check of the  $\mathcal{O}$(5\%) HEU deficit recently observed at $\sim$10~m by STEREO~\cite{stereo_rate}.   

Measurement of the pure $^{235}$U $\overline{\nu}_e$ flux with the existing run of PROSPECT data is challenging, given the complexity of precisely calibrating IBD positron energy loss and IBD neutron migration effects in the vicinity of non-functioning detector segments. 
This can be partially mitigated by restricting the IBD event selection to the highest functioning region of the detector, but statistical uncertainties from such a selection are expected to be between 1-2\% for the first run of PROSPECT.  
In a PROSPECT-II detector with all segments functioning, target-external radioactive and neutron source deployment systems will enable calibration of total IBD detection efficiency to well below percent-level precision.  
Statistical uncertainties generated by expected signal and background in two years of PROSPECT-II operation at HFIR, given in Table~\ref{tab:exp_param}, are likely to be at the 0.3\% level or lower.  
Combined with 2\%-level knowledge of HFIR's average thermal power using its current monitoring instrumentation~\cite{conant_thesis}, a $\sim$2.5\% precision measurement of the $^{235}$U $\overline{\nu}_e$ flux should be achievable with PROSPECT-II.  

\begin{table*}[thbp!]
\centering
\begin{tabular}{c|c||c|c|c}
\hline
\multirow{2}{*}{Case} & \multirow{2}{*}{Description} & \multicolumn{3}{|c}{Precision on $\sigma_i$ (\%)} \\ \cline{3-5}
& & \uFive &\pNine &\uEight \\ \hline \hline
1& Daya Bay LEU & 3.7 & 8.2 & 30  \\ \hline
2& Daya Bay LEU + P-II HEU & 2.4 & 6.3 & 21.3  \\  \hline
3& P-II LEU + P-II HEU+  & 1.4 & 3.4 & 15.9  \\ \hline
4& P-II LEU + P-II HEU+, Correlated & 1.4 & 3.0 & 8.7  \\ \hline\hline
- & Model Uncertainty~\cite{huber_berryman} & 2.1 & 2.5 & 11.2  \\ \hline
\end{tabular}
\caption{Constraints on IBD yields of \uFive, \pNine, and \uEight~from future hypothetical datasets from LEU and HEU reactors, given as a percentage of the best fit yield.  
'P-II' refers to PROSPECT-II, 'HEU+' refers to a HEU-based measurement with thermal power uncertainty improved from 2\% to 1\%, and 'Correlated' refers to correlated detector systematics between HEU and LEU measurements.  
For all cases, a 30\% constraint is used for \uEight, roughly matching current levels of disagreement between global fits and summation models~\cite{giunti_diagnose}.  
Model prediction uncertainties from ~\cite{huber_berryman} are also provided for reference.  
}
\label{tab:Uncertainty_Improvement}
\end{table*}

Table~\ref{tab:Uncertainty_Improvement} overviews results from combined fits of various existing and future hypothetical PROSPECT-II flux datasets used to determine expected uncertainties on best-fit isotopic IBD yields for \uFive, \pNine, and \uEight.  
Fit procedures mirror those previously described in Ref~\cite{surukuchi_flux}; the assumed uncertainty constraint on the flux of the sub-dominant isotope \pOne{} is 2.2\%, as in Ref.~\cite{bib:huber}, while \uEight~uncertainties are set to 30\%, roughly matching current levels of disagreement between global fits and summation models~\cite{giunti_diagnose}.  
This looser \uEight~constraint enables study of how \nuebar flux measurements might provide independent knowledge regarding the accuracy of nuclear data for this fission isotope.  
No constraints are applied to \uFive~and \pNine~IBD yield fit parameters, and sterile oscillation effects are also not included in the fit ($\theta_{14}=0$).  

When combined with Daya Bay flux results~\cite{bib:prd_rate}, it is clear that a 2.5\% precision IBD yield measurement from PROSPECT-II can also yield improved knowledge of individual isotopic yields for the other fission isotopes \pNine~and \uEight.  
When adding PROSPECT-II to a Daya Bay yield fit (first versus second row in Table~\ref{tab:Uncertainty_Improvement}), \uFive~and~\pNine~uncertainties are improved by a relative margin of 50\% and 30\%, respectively, and direct constraints on \uEight~yields are enabled.  
A future program of dedicated HFIR thermal power calibration using neutron activation analysis (NAA) of in-core samples may enable further improvements in power measurement uncertainty to $\sim$1\% precision by addressing its dominant source of systematic uncertainty, possibly enabling a future PROSPECT-II \uFive~IBD yield measurement precision of better than 1.5\%.  
The collaboration has initiated R\&D regarding NAA-based estimates of integrated thermal power.

\subsection{Physics goals of subsequent reactor deployments}
\label{subsec:phys_mult}

Beyond the program at HFIR outlined above, subsequent redeployment of PROSPECT-II at other reactor locations will be possible with the improved robustness of the PROSPECT-II detector design.  
A natural choice for future deployment would be a commercial LEU reactor core.  
Successful US-based short-baseline commercial core deployments have been performed recently both on the earth's surface~\cite{Haghighat:2018mve} and underground~\cite{SONGS1,bowdenmon}.  
Additional reactor-based redeployments that could be considered for PROSPECT-II include fast reactors burning mixed oxide (MOX) fuel~\cite{bernmon2,huber_mox,ismran}, such as the planned US-based Versatile Test Reactor (VTR)~\cite{aap2018}. The fission fractions in a MOX reactor are very different from an HEU or LEU reactor; for example, estimated fission fractions for the VTR are 62\% \pNine{}, 8\% $^{240}$Pu, 4\% \pOne{}, 13\% \uFive, and 13\% \uEight.  

The physics benefits of an LEU-based deployment are explored using the experimental parameters presented in the rightmost column of Table~\ref{tab:exp_param}.  
These reactor power and baseline parameters roughly match those realized at previous commercial reactor deployments.  
Reactor-on and reactor-off days match those achievable in a measurement over a full 18-month fuel cycle plus two month-long refueling periods before and after the cycle.  
The signal-to-background ratio of 19.3 is calculated assuming an on-surface deployment with a  background rate comparable to that of the HFIR deployment and a signal rate increased due to the higher LEU reactor power.  
As indicated by the relative difference between true and effective IBD statistics for the LEU case, modest gains beyond those presented below may be achieved through underground deployment.  

The sterile neutrino oscillation sensitivity benefits provided by an LEU-based deployment of PROSPECT-II, following the HFIR deployment, are illustrated in the right panel of Figure~\ref{fig:osc}.  
The LEU deployment provides a new high-statistics dataset at longer baselines, which extends the PROSPECT-II region of peak sensitivity from  $\sim$1-20~eV$^2$ for an HEU-only deployment all the way to the sub-0.3~eV$^2$ regime dominated by Daya Bay.  
When LEU data-taking is followed by a return to HFIR for two years of additional data-taking, PROSPECT-II and other reactor-based efforts can ensure that the sin$^22\theta_{14}=0.03$ benchmark for clarifying long-baseline interpretations is met in the \nuebar disappearance channel over the entire regime from $\Delta$m$^2_{32}$ to 10~eV$^2$.  

Deploying PROSPECT-II at an LEU reactor after HFIR would also expand the reactor \nuebar spectrum physics.  
PROSPECT measurements at both HEU and LEU reactors will enable one of the most precise determinations yet of \nuebar spectra from both \uFive{} and \pNine{}.  
Such a measurement has advantages over a joint analysis of PROSPECT HEU data with an existing LEU dataset, such as Daya Bay, in two key respects.  
Since a short-baseline LEU deployment samples \nuebar flux from a single core, as opposed to multiple cores of varying fuel content, it is better able to observe spectra produced at very high and very low \pNine~content, enabling a more precise untangling of \uFive~ and \pNine~spectrum contributions. NEOS-II plans to produce such an LEU measurement~\cite{neos2}.  
In addition, detector systematics contributing to HEU and LEU spectrum uncertainties will be highly correlated when both are measured with the same PROSPECT-II detector, which will enhance \uFive~and \pNine~spectrum precision and allow measurement of~\uEight~contributions not present in the HEU observation.  
Apart from LEU data-taking, subsequent redeployment at HFIR for another two years of data-taking (four calendar years of total HEU data-taking) would result in a fully systematics-dominated \uFive~measurement.  
Deployment at a MOX reactor offers other spectrum measurement opportunities. Due to the very different relative contribution of plutonium fissions with respect to LEU or HEU cores, a MOX-based PROSPECT-II measurement would improve direct understanding of  \nuebar spectra produced by \pNine, \pOne, and $^{240}$Pu.

As with spectrum and oscillation measurements, redeployment of PROSPECT-II at an LEU reactor following a two-year HFIR deployment would enhance the scope of PROSPECT's \nuebar flux physics capabilities.  
The gains of a combined LEU-HEU PROSPECT flux measurement  are illustrated in the third and fourth row of  Table~\ref{tab:Uncertainty_Improvement}.  
Both rows assume HEU thermal power measurement uncertainties of 1\%, which, as described in Section~\ref{subsec:phys_flux}, may be achievable through further R\&D.  
Combined HEU and LEU measurements in a scenario of fully uncorrelated systematics (row 3) yield 40\%, 45\%, and 25\% improvement in IBD yield precision for \uFive, \pNine, and \uEight~relative to a scenario involving a combination of PROSPECT-II HEU and Daya Bay LEU measurements (row 2).  
If fully correlated systematic uncertainties can be demonstrated between HEU and LEU deployments of PROSPECT-II (row 4), direct \nuebar-based constraints on isotopic fluxes will rival or exceed the precision of state-of-the-art models for all three of these primary fission isotopes.  
MOX-based flux measurements with PROSPECT-II would provide additional understanding of integral \nuebar production by multiple plutonium isotopes.

%% file: PI_Detector.tex
PROSPECT-II will be an optimized evolution of the PROSPECT-I detector.  To provide context, the following sections describe relevant aspects of the design and performance of PROSPECT-I.
The active detector is a segmented 4-ton $^6$Li-doped liquid scintillator (LiLS) volume. 
Antineutrinos with an energy above the 1.8 MeV threshold are detected via the
inverse beta-decay (IBD) reaction on protons in the LiLS.
The positron, which receives most of the antineutrino energy above the reaction threshold, rapidly
annihilates with an electron to produce a prompt signal with energy ranging from 1 MeV to $\sim$8 MeV. 
The neutron captures on $^6$Li or H with a typical capture
time of 40\,$\mu$s.  
Neutron captures on $^6$Li (denoted `nLi') produce well localized energy depositions from the reaction $n$ + $^6$Li$ \rightarrow{} \alpha + t + 4.8$ MeV. The quenched signal from these products amounts to about 0.55 MeV. 
As this capture only produces heavy charged particles, the Pulse Shape Discrimination (PSD) capabilities of the LiLS are able to separate neutron captures from background  $\gamma$-ray events, reducing random coincidences~\cite{prospect_p20}. 
More importantly, the correlation in time and space between the prompt and delayed signals, both with excellent PSD discrimination, provides a distinctive antineutrino signature, greatly suppressing backgrounds.  PROSPECT is unique in fully leveraging both prompt and delayed PSD. 
Since the LiLS and segment design enable a precise measurement of prompt energy, and the detector segments are between 7-9 meters from the compact HFIR core, the \nuebar{} interaction probability is measured as a function of both baseline and energy, enabling a highly selective search for neutrino oscillations in addition to a high-resolution spectrum measurement. 

\begin{figure}
\centering
\includegraphics[trim=0mm 50mm 0mm 60mm, width=0.85\textwidth]{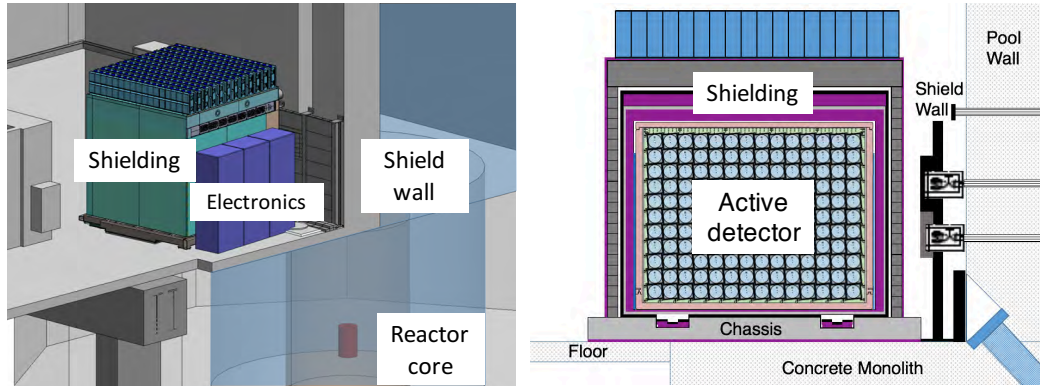}
\caption{
Left: Layout of the PROSPECT experiment, with 
the detector installed in the HFIR Experiment Room next to the water pool. The detector is 5.09\,m above the reactor core (dark red). The distance from the center of the reactor core to the center of the detector is 7.84 m.  (Right) Schematic showing the active volume divided into 14 (long) by 11 (tall) segments and surrounded by containment vessels  and shielding layers.
A shield wall covers penetrations in the pool wall associated with high backgrounds.
}
\label{fig-layout}
\end{figure}

The PROSPECT-I detector is shown in Figure~\ref{fig-layout} 
as installed at HFIR next to the reactor water pool.
The detector consists of the LiLS inner volume, an acrylic inner containment vessel (ICV), an outer aluminum containment vessel (OCV), shielding, detector  movement
elements, and data acquisition (DAQ) and control electronics~\cite{prospect_nim}.
The planned PROSPECT-II upgrade makes significant modifications only inside of the OCV.  In PROSPECT-I, the active LS volume is divided into 14 by 11 segments  by reflective optical separators held together at the edges by 3D printed hollow `corner rods'~\cite{prospect_grid}.
Segments are parallel to the reactor pool wall on the north side of the detector and approximately perpendicular to the antineutrino direction.
Each segment is viewed on the east and west ends by 5 inch PMTs enclosed in mineral oil-filled acrylic housings.
Acrylic supports tie the housings together and support the outermost optical separators and corner rods.
The detector was transported from its assembly site to ORNL while dry and filled onsite. 
The top layer of optical separators is covered by a few cm of LiLS, with an expansion volume above containing nitrogen cover gas.  

The inner detector has several unique design features:
\begin{itemize}
\item{Liquid scintillator doped with 0.08\% $^6$Li by mass, achieved via a reverse micelle phase containing LiCl.  This provides a  localized neutron capture which is easily separated from $\gamma$-ray backgrounds 
through topology and PSD.
}
\item{A reflective grid separating the active volume into 154 segments of uniform volume, providing event localization and enhanced background rejection.
Neighboring segments share optical separators made of an opaque low-mass carbon fiber core covered by laminated reflective and fluorinated ethylene propylene (FEP) film. 
}
\end{itemize}

\begin{itemize}
\item{A tessellated segment structure that minimizes non-reflective surfaces in the optical volume while providing access for multiple optical or radioactive calibration sources throughout the active detector volume. }
\item{PMT housings protruding slightly into the optical grid, minimizing cross-talk and improving the optical coupling to the segment interior.
}
\end{itemize}

The LiLS is chemically incompatible with a broad range of materials, so materials exposed to it were extensively tested and limited to cast and machined acrylic surfaces; polytetrafluoroethylene (PTFE); fluorinated ethylene propylene (FEP) film; Viton O-rings; polyether ether ketone (PEEK) screws; 3D printed polylactic acid (PLA) and the nitrogen cover gas.

\noindent
The visible LiLS volume in each PROSPECT-I segment is 
14.4~cm~(wide)  by 14.4~cm~(tall)  by 118~cm~(long).
The housings support the corner rods and define the segment geometry.
Selected  rods contain tubes for the insertion of radioactive sources into the active volume or optical diffusers midway along the segment length that are coupled to an optical calibration system. 
The PMT housings, immersed in the common LiLS volume, contain the PMT and divider, reflectors, and magnetic shield.  

PROSPECT's precision goals were unprecedented for an on-surface segmented detector; thus great care was taken to achieve a well-understood detector response.
This led to two primary design features: (1) insertion of PMT housings into the optical reflector grid to eliminate cross-talk, and (2) a full-volume calibration system~\cite{prospect_cal}. 
The combination of these features necessitated incorporation of PMT housings within the LiLS volume.  
However, during \Pone{} operation, LiLS was observed to have entered the PMT housings, leading to failures of the PMT electronics and the loss of those segments as active target region. 
Additionally, the large number of required penetrations of the inner acrylic tank combined with the diversity of materials in contact with the LiLS likely led to the slow degradation of scintillator light yield. 
In light of the lessons learned from \Pone{}, these design choices are reconsidered for PROSPECT-II.  

%% file: PII_Detector.tex
\begin{figure*}
\centering
\includegraphics[clip=true, trim=0mm 0mm 0mm 45mm,width=0.75\textwidth]{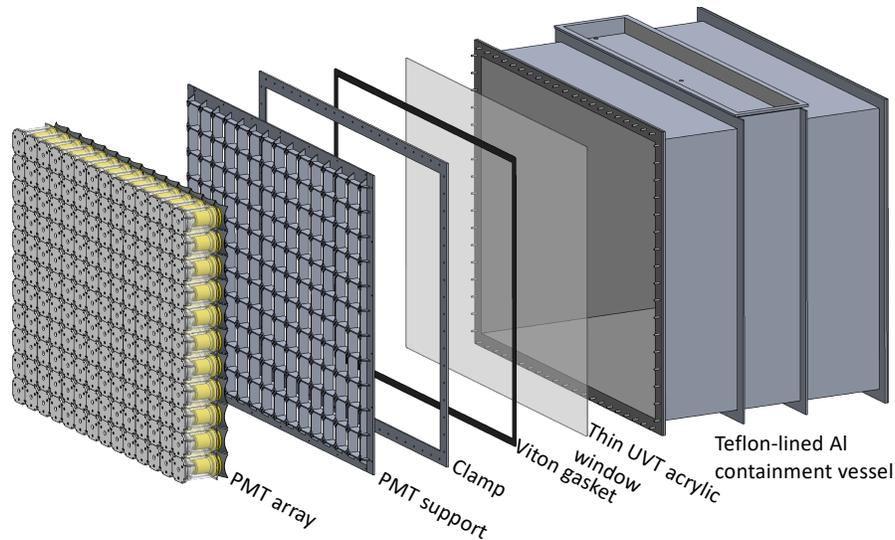}
\caption{Schematic drawing of the redesigned inner detector. Arrays of 5" PMTs are mounted in aluminum support structures
on both ends of the inner containment vessel (for simplicity, this drawing shows only the PMTs and mounts on one side). The support structure also
reinforces the thin acrylic window between the LiLS volume and PMTs. Mineral oil fills the spaces between the inner detector and outer containment vessel (not shown). }
\label{fig:PII}
\end{figure*}

The PROSPECT collaboration has developed an evolutionary design for a detector upgrade that addresses the technical issues encountered during operation of \Pone{}, requires modification to only a minority of subsystems, and maintains the performance required to achieve the physics goals of the experiment.  
As mentioned above, the planned PROSPECT-II upgrade makes significant updates to the original PROSPECT design only inside of the OCV.  
The principal change to the inner detector design moves the PMTs outside the LiLS volume to reduce the possibility of PMT voltage divider degradation and reduces the range of materials in contact with LiLS, thereby providing a suitable environment for long-term stability.
Executing the upgrade involves a rebuild of the inner detector and production of new LiLS with a slightly higher $^6$Li loading fraction, leaving the OCV, shielding package and DAQ largely unchanged.  

\subsubsection{Containment vessel}

The centerpiece of the PROSPECT-II upgrade is therefore the inner detector shown in Figure~\ref{fig:PII}. 
Existing PMTs, outer aluminum OCV, shielding, support structure and electronics from the original PROSPECT will be reused with only minor modifications. 
Importantly, the upgrade concept keeps \Ptwo{} within the established size and weight footprint approved for operation at HFIR, minimizing the effort needed for project approval and detector installation.

A PTFE-lined aluminum inner containment vessel (ICV) contains the target LiLS and optical reflector panels. 
Thin UVT acrylic windows on each end provide the optical interface from the LiLS target to the PMTs.
Double O-rings between the window and the vessel will allow verification and monitoring of the seal.  
An aluminum frame on the exterior of the ICV provides mechanical support for the interface window and  mounting points for the PMTs that match the segment locations within. 
The space between the inner and outer containment vessels is filled with mineral oil, providing optical coupling and reducing backgrounds.
This design allows for full control of the nitrogen LiLS cover gas, minimizes the types of materials exposed to the LiLS, and  significantly reduces the area of surfaces exposed to the LiLS, thus decreasing the risk of LiLS performance degradation over the planned operation period.  

\subsubsection{Inner detector structure}
The inner detector is structurally robust, allowing transportation by commercial air-ride trailer, and the LiLS can be drained and refilled while maintaining control of the target region atmosphere.
These features  enable the LiLS to be exchanged or replaced \textit{in situ} if needed and permit a relatively straightforward redeployment at another reactor site to expand the PROSPECT-II physics program.

\begin{figure}
\centering
\includegraphics[clip=true, trim=5mm 0mm 150mm 110mm,width=0.47\textwidth]{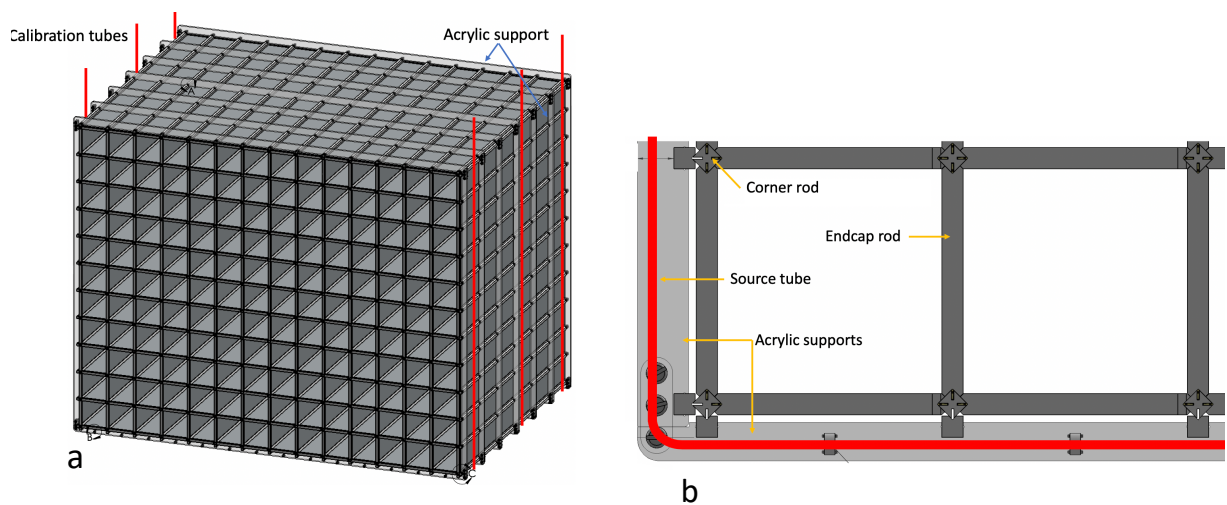}
\includegraphics[clip=true, trim=130mm 10mm 10mm 120mm,width=0.5\textwidth]{figures/Fig8.pdf}
\caption{ Left: Placement of the calibration tubes (red) along the perimeter of the segment array. 
Right: End view showing the endcap rods and corner rods supporting the reflector panels. 
Panels are spaced away from the containment wall by acrylic supports to allow space for fill lines, level sensors, and calibration tubes (red). 
}
\label{fig:Grid}
\end{figure}

As in \Pone{}, the reflectors are flat panels of laminated FEP film, DF2000MA reflector, and carbon fiber. 
The reflecting panels are held in place by corner rods machined from PTFE bar stock. 
Materials exposed to LiLS in this design are limited to FEP, PTFE, Viton, and  cast UVT acrylic, avoiding large areas of machined acrylic which were difficult to clean in the \Pone{} design.
The outermost reflector panels are separated from the inner containment wall by acrylic spacers, leaving space for fill/drain lines, level sensors, and PTFE calibration source tubes, as shown in Figure~\ref{fig:Grid}. 
Additional PTFE pieces on the ends of the panel ensure a tight fit against the acrylic windows, maintain the segment geometry, and allow liquid and gas to flow between segments via angled slots.
The thickness of the acrylic window is kept to $\sim$5 mm to minimize cross-talk between neighboring segments.
The LiLS and mineral oil fluid levels will be roughly matched during filling or draining to reduce stress on the window from hydro-static pressure. 

Removing the PMTs from the LiLS volume requires a new structure to mechanically support the PMTs, provide optical separation, and increase rigidity of the acrylic window, as shown in Figure~\ref{fig:PII}.
Constructed from aluminum, the rectangular grid separates the PMTs and provides a mounting point for the individual PMT supports, including a conical front-reflector and magnetic shielding. 

\subsubsection{Detector response}
\label{sec:det_response}

\begin{figure}
  \centering
\includegraphics[clip=true, trim=0mm 0mm 0mm 00mm,width=0.32\textwidth]{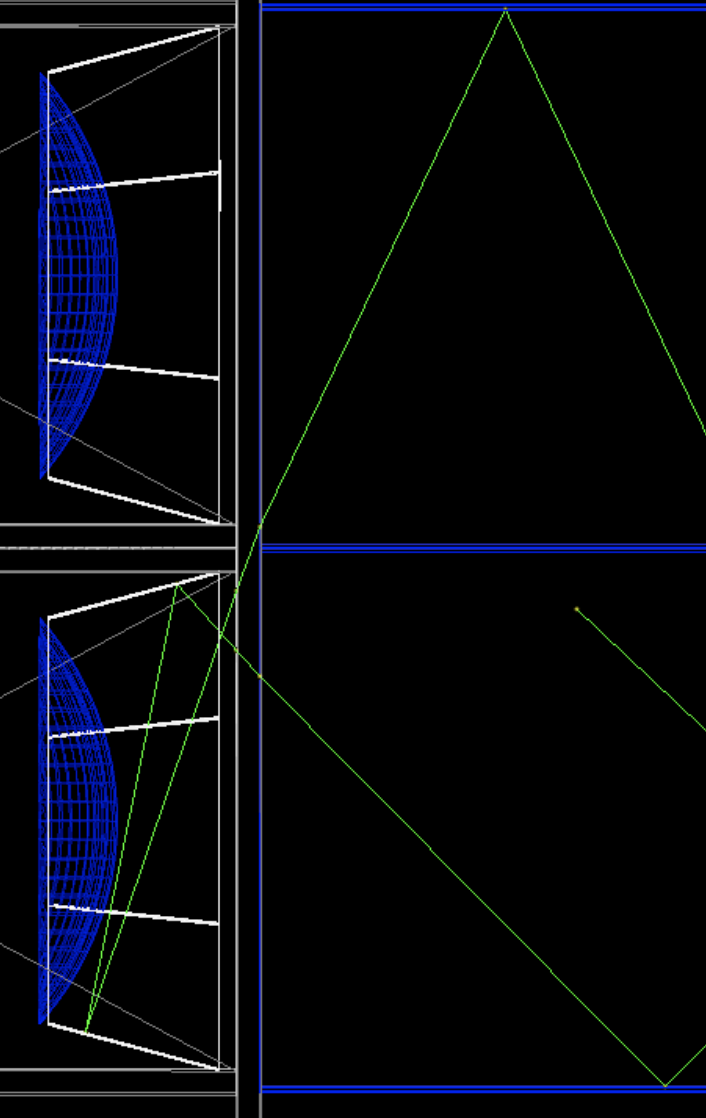}
\includegraphics[clip=true, trim=0mm 5mm 15mm 10mm,width=0.55\textwidth]{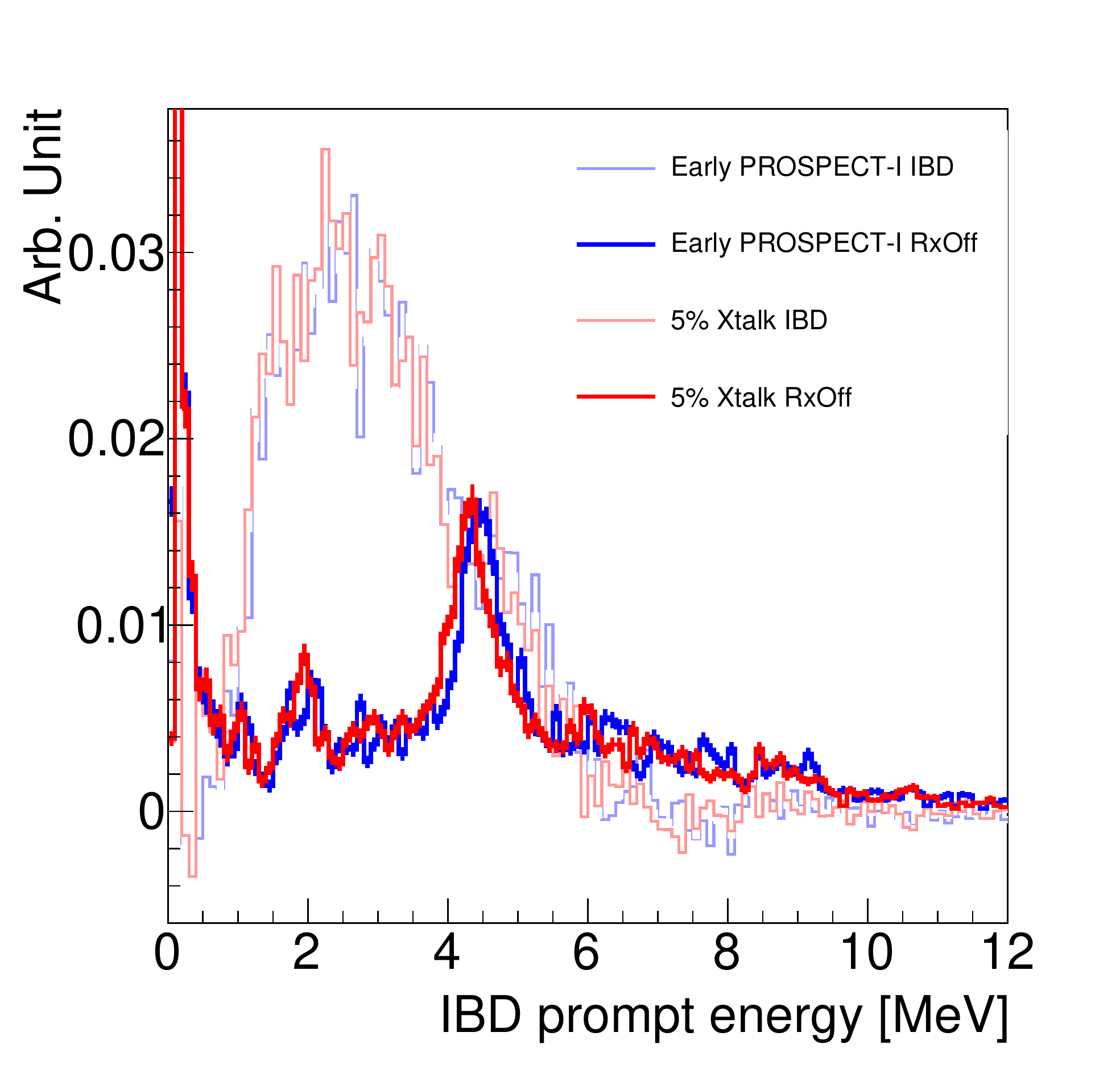}
\caption{ Left: GEANT4 rendering of cross-talk mechanisms in PROSPECT-II.  An optical photon (green) is shown propagating from one PROSPECT-II segment into another; blue horizontal lines, white trapezoids, and blue parabolic shapes represent reflective segment walls, PMT reflector cones, and PMT surfaces, respectively.  Right: Measured \Pone{} IBD-like prompt spectra before (red) and after (blue) introducing artificial cross-talk.  Datasets are shown for reactor-off periods, as well as for reactor-on periods after applying full background subtraction.}
\label{fig:crosstalk}
\end{figure}

The PROSPECT-II upgrade introduces three significant changes that impact the detector response. 
First, the single optical interface window between the LiLS target and the PMTs will allow a small amount of optical cross-talk between segments.  
Optical simulations have illustrated primary cross-talk pathways, as shown in Figure~\ref{fig:crosstalk}, and provide evidence that percent-level cross-talk will be present in the nominal PROSPECT-II detector design.  

The impact of this cross-talk on the event reconstruction ability of \Ptwo{} has been evaluated using data from the first month of \Pone{} operation.
Energy reconstruction is not of primary concern, since reconstructed prompt energy is the sum over contiguous modules above threshold. 
Instead, the focus is on the ability to identify and reject background, since cross-talk from a high-energy electromagnetic deposition in one segment could mask a low-amplitude neutron recoil event in an adjacent segment, where the identification of the latter via PSD is important for background rejection.  
Cross-talk was artificially introduced to collected \Pone{}~data by applying a response function that distributes a fraction of event signal amplitude in one segment to its neighbours. 
As shown in Figure~\ref{fig:crosstalk}, a cross-talk magnitude of $\sim$5\% is found to modestly decreases the signal-to-background ratio in the examined dataset from 2.9:1 to 2.8:1.
This data-based analysis demonstrates that cross-talk magnitudes expected in the \Ptwo{}  will not substantially affect IBD event selection capabilities.

Second, full-volume calibration access to the target mass will be substantially reduced, if not entirely eliminated. 
The PROSPECT-II calibration strategy avoids many penetrations into the ICV and simplifies design of the optical grid. 
Calibration access tubes will be run around the periphery of the ICV at several locations as shown in Figure~\ref{fig:Grid}.
Monte Carlo calculations for PROSPECT-I showed excellent agreement with data throughout the detector volume, including source positions at the detector edges, where energy leakage is significant~\cite{prospect_prd}.  
Accordingly, the PROSPECT-II detector energy scale will be established by deploying several $\gamma$-ray sources at positions around the edge of the active volume in conjunction with the beta spectrum of cosmogenically produced $^{12}$B.  

\begin{figure}
  \vspace{-8mm}
\centering
\includegraphics[clip=true, trim=0mm 88mm 0mm 0mm,width=0.98\textwidth]{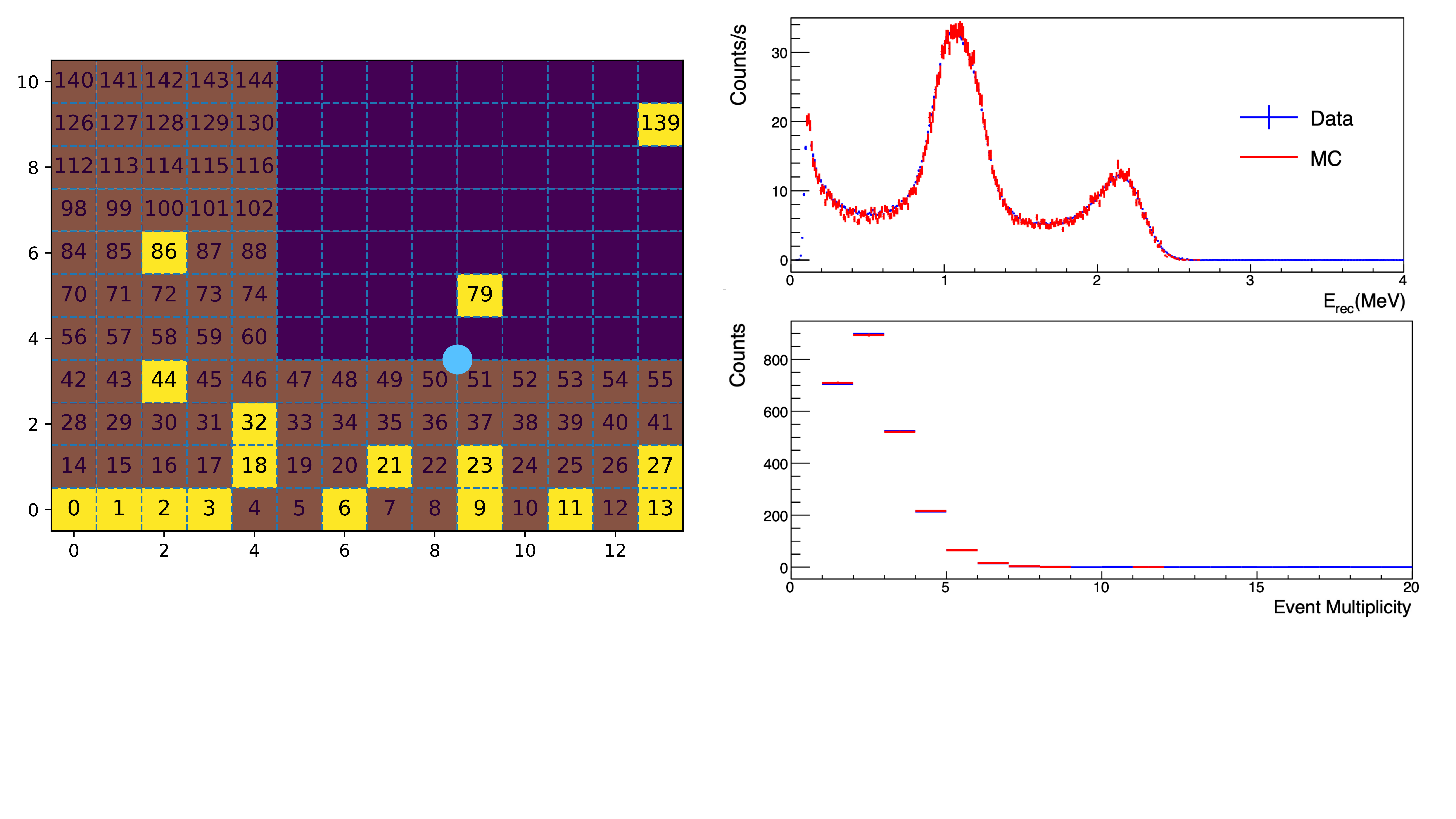}
\caption{Left: Illustration of PROSPECT-I segments analyzed to mimic externally-deployed calibration source datasets.  Dark blue, brown, and yellow squares represent analyzed, un-analyzed, and inoperative segments, while the light blue circle represents the deployed $^{60}$Co source location.  Right: Simulation-data agreement for energy (top) and multiplicity (bottom) distributions from dark blue segments.}
\label{fig:off-center-cal}
\end{figure}

To confirm this approach, \Pone{} data has been analyzed to mimic the proposed \Ptwo{} source deployment configuration, as shown in Figure~\ref{fig:off-center-cal}.  
In this case, only half of the segment data from a centrally deployed $^{60}$Co source is analyzed to mimic an external location. 
Following the procedure used in \Pone{}, the energy scale is established using both energy and multiplicity spectra. 
For this approach to be accurate in \Ptwo{}, detector components outside the active portion of the ICV will need to be properly represented in the detector simulation package.
In addition, experience with PROSPECT-I has also shown that drifts in detector calibration and performance can be adequately tracked using intrinsic sources such as thermal neutron captures on hydrogen. 
Likewise, timing information from the optical calibration system can be replaced by information from through-going cosmic ray muons, already used for some timing information in PROSPECT-I. 

Finally, the more space-efficient PMT support structure allows for elongation of the active target region. 
This increases the detector active volume by $\sim$20\%, which corresponds to a 30\% increase in fiducial volume. 
Using the validated GEANT4-based PROSPECT detector simulation package PG4~\cite{prospect_prd}, the IBD and background detection rates for the \Ptwo{} detector can be estimated.  
As previously shown in Table~\ref{tab:exp_param}, both performance metrics exceed those demonstrated in \Pone{}, due to the absence of inoperative segments in PROSPECT-II.  
With a segment length equal to that in \Pone{} (118~cm), an IBD event rate of roughly $900$/day is expected with a signal-to-background ratio of 3.9. 
With the larger segment length of $145$~cm, the IBD rate increases to roughly 1150/day with a signal-to-background ratio of 4.3.

\subsubsection{$^6$Li-loaded liquid scintillator}
\label{sec:LiLS}

New liquid scintillator with a $^6$Li-loading of $0.1\%$ by mass (increased from $0.08\%$) will be produced for \Ptwo{}.
The continued use of the present LiLS formulation is well supported by the light yield achieved and excellent signal-to-background and PSD observed with \Pone{}.
Available evidence suggests that the declining light collection observed in the first run of PROSPECT is an environmental effect from oxygen quenching or material incompatibility rather than an intrinsic property of the LiLS.  
Measurements to confirm this experimentally are currently underway.  
A near-term R\&D program has been developed to quantitatively assess a variety of LiLS samples that have  been be in storage for several years. 

Measurements of these LiLS samples to date have not shown a significant degradation of the UV-absorption characteristics consistent with that observed during \Pone{} operation.  
Previous bench testing has shown that the observed changes could be consistent with either oxygen quenching or yellowing due to material incompatibility or some combination of the two.  
To explore possible causes, samples of LiLS drawn from the \Pone{} detector will be analyzed 
before and after nitrogen sparging using bench-top UV-vis spectrometry and coincidence gamma counting.
Studies of micelle stability and light scattering, which could modify the effective attenuation length, will also be performed.
Such tests will allow a disentangling of the intrinsic stability, material interaction (detector components), and operating environment (nitrogen buffer gas).  
Similar testing will be carried out on a new panel of months-long soak tests of material samples of inner materials from \Pone{}.  
Such testing can address potential variations in material production runs and changes in behavior due to machining.

\subsubsection{Infrastructure improvements}

Beyond the significant design upgrades in the OCV, minor upgrades are anticipated for other systems. Weight and size constraints preclude major changes to the \Pone{} shielding design.
A rearrangement of shielding in forklift channels will reduce the only observed hot spot in the reactor-on gamma backgrounds by $\approx 50$\%, thereby reducing DAQ dead-time and accidental backgrounds. 
Modest additions to thermal neutron shielding may reduce the flux of neutron-capture $\gamma$-rays from building material surrounding the detector.

The new inner containment vessel design separates the nitrogen cover gas volume over the LiLS from the gas over the mineral oil surrounding the PMTs.
This ensures better control of the humidity and oxygen contamination of the LiLS cover gas.
The number of penetrations is significantly reduced, greatly simplifying the leakage requirements of various seals and feedthroughs.

The temperature of the LiLS and surrounding buffer was not actively controlled in PROSPECT-I. 
Turning on the PMTs raised the LiLS temperature by 2-3 $^\circ$C. 
Larger LiLS temperatures excursions of  $\sim~6 ^\circ$C were observed during occasional multi-day chiller outages  at HFIR. 
The PROSPECT-II detector will be actively cooled by adding a cooling loop to the mineral oil and an external heat exchange. 
The coolant temperature can be maintained at a desired set point and should limit changes to $\sim 1^\circ$C.

%% file: Conclusions.tex
The PROSPECT program has, in its first phase, demonstrated the ability to detect reactor antineutrinos in an aboveground detector, set new limits on active-sterile neutrino mixing, and produced the world's most precise measurement of antineutrinos from an HEU reactor. These contributions represent important progress toward resolving the Reactor Antineutrino Anomaly and achieving more complete knowledge of reactors as an antineutrino source for basic science and possible applications. However, important goals remain to be reached: a region of sterile neutrino phase space that could explain the anomaly remains unchecked, and unexpected features in the reactor antineutrino spectrum remain to be clarified. 

An evolutionary upgrade of the PROSPECT detector, PROSPECT-II, is now under development to reach these goals in a longer deployment at the HFIR reactor. Principle upgrades are moving the PMTs out of the scintillator volume, reducing materials in contact with the scintillator, and increasing the detector volume and $^{6}$Li doping fraction, which together allow a longer data-taking run and an order-of-magnitude increase in effective signal statistics. With this new dataset, PROSPECT-II will cover a previously unexplored region of sterile neutrino phase space in the 1-20 eV$^2$ mass splitting range (see Figure \ref{fig:osc}). Addressing this region will be an important component in conclusively resolving the Reactor Antineutrino Anomaly. This electron-flavor disappearance search is complementary to searches for new physics with muon-flavor sources.

The reach of PROSPECT-II will extend to sterile neutrino possibilities beyond the Reactor Antineutrino Anomaly, with relevance to the interpretation of long-baseline CP violation measurements. In particular, PROSPECT-II will extend global sensitivity to the sterile mixing angle $\theta_{14}$ below $\sim$5$^{\circ}$ in the mass splitting region of $\sim1-10$ ev$^2$ (see Figures \ref{fig:current_osc} and \ref{fig:osc}. As discussed above, reaching that benchmark eliminates certain complications in the interpretation of CP violation searches.

Finally, PROSPECT-II will reduced uncertainties on the measured $^{235}$U antineutrino spectrum uncertainties below 5\%, providing a unique constraint on reactor antineutrino models (see Figure \ref{fig:spectrum}). An additional deployment at a commercial nuclear reactor would expand upon these contributions to the particle physics, nuclear science, and nuclear security communities.